\documentclass[hyper,letterpaper,notoc]{JHEP3}
\pdfoutput=1
\usepackage{amsmath}
\usepackage{graphics}
\usepackage[config=altsf]{subfig}
\usepackage{feynmf}
\usepackage{dsfont}
\usepackage{array}
\usepackage{slashed}
\usepackage{afterpage}
\usepackage{appendix}
\usepackage{multirow}

\newcommand{\wt}{\widetilde}

\title{A Classification of Dark Matter Candidates with Primarily 
Spin-Dependent
Interactions with Matter}
\author{Prateek Agrawal\\ Maryland Center for Fundamental Physics,
Department of Physics, \\University of Maryland,
College Park, MD 20742}
\author{Zackaria Chacko\\ Maryland Center for Fundamental Physics,
Department of Physics, \\University of Maryland,
College Park, MD 20742}
\author{Can Kilic\\ Department of Physics and Astronomy, \\Rutgers University,
Piscataway NJ 08854}
\author{Rashmish K. Mishra\\ Maryland Center for Fundamental Physics,
Department of Physics, \\University of Maryland,
College Park, MD 20742}

\abstract {We perform a model-independent classification of Weakly Interacting 
Massive Particle (WIMP) dark matter candidates that have the property that their 
scattering off nucleons is dominated by spin-dependent interactions. We study 
renormalizable theories where the scattering of dark matter is elastic and arises 
at tree-level. We show that if the WIMP-nucleon cross section is dominated by 
spin-dependent interactions the natural dark matter candidates are either 
Majorana fermions or real vector bosons, so that the dark matter particle is its 
own anti-particle. In such a scenario, scalar dark matter is disfavored. Dirac 
fermion and complex vector boson dark matter are also disfavored, except for very 
specific choices of quantum numbers. We further establish that any such theory 
must contain either new particles close to the weak scale with Standard Model 
quantum numbers, or alternatively, a $Z'$ gauge boson with mass at or below the 
TeV scale. In the region of parameter space that is of interest to current direct 
detection experiments, these particles naturally lie in a mass range that is 
kinematically accessible to the Large Hadron Collider (LHC).}

\preprint{UMD-PP-10-004\\RUNHETC-2010-07}

\begin{document}
\section{Introduction}

While it is now well established that about 80\% of the matter in the 
universe is non-luminous and non-baryonic, the exact nature of the 
particles that constitute this dark matter remains a mystery.  It is 
widely known, however, that weak scale stable particles that interact with 
visible matter with strength comparable to the weak force naturally tend 
to have the right relic abundance to explain observations. Then perhaps 
the simplest candidates for dark matter are such Weakly Interacting 
Massive Particles (`WIMP's). Many well-motivated extensions of the 
Standard Model (SM) contain WIMP dark matter candidates that have been 
shown to yield the correct relic abundance, for example 
supersymmetry~\cite{Ellis:1983ew,Griest:1988ma}, extra dimensional 
theories~\cite{Kolb:1983fm,Servant:2002aq}, little Higgs 
models~\cite{BirkedalHansen:2003mpa,Hubisz:2004ft} and the left-right twin 
Higgs model~\cite{Dolle:2007ce}.

A variety of current and future experiments are involved in the search for 
WIMP dark matter. These include direct detection experiments, which aim to 
directly observe dark matter, indirect detection experiments, which search 
for the annihilation products of dark matter, and collider experiments, 
which hope to directly produce dark matter. The signal in current direct 
detection experiments is the kinetic energy transferred to a nucleus after 
it scatters off a dark matter particle. The energies involved are less 
than or of order 10 keV, which is well below the typical nuclear energy 
scales. Therefore, at these energies, the WIMP sees the entire nucleus as 
a single unit, with a net mass, charge and spin.

The magnitude of the WIMP-nucleus scattering cross section is extremely sensitive 
to the exact form of the interactions of the dark matter particle with the 
individual nucleons. In theories where the WIMP couples primarily to the spin of 
the nucleon, the corresponding interactions are labelled as spin-dependent. On the 
other hand, in theories where the WIMP-nucleon cross section is insensitive to the 
spin of the nucleon, the corresponding interactions are labelled as 
spin-independent. For a given WIMP-nucleon cross section, the corresponding 
WIMP-nucleus cross section is in general significantly larger for spin-independent 
interactions than for spin-dependent interactions. Why is this?  Spin-independent 
scattering tends to be coherent, receiving contributions from all the nucleons in 
the nucleus. The corresponding WIMP-nucleus cross section is enhanced by a factor 
of $A^2$, where $A$ is the mass number of the nucleus. On the other hand, since the 
spins of nucleons in a nucleus tend to cancel in pairs, there is no such 
enhancement in the spin-dependent WIMP-nucleus cross section.  For this reason, 
direct detection experiments are much more sensitive to spin-independent 
interactions than to spin-dependent interactions. In particular, the current bound 
on spin-independent WIMP-nucleon interactions is about five orders of magnitude 
stronger than the corresponding bound on spin-dependent interactions 
\cite{Ahmed:2009zw,Aprile:2009zzb,Angle:2008we,Kim:2008zzn}.

Clearly, the prospects for direct detection of dark matter depend crucially on 
whether WIMP-nucleon interactions are primarily spin-dependent or spin-independent. 
Recently, several authors have studied the relative sizes of the spin-dependent and 
spin-independent contributions to the WIMP-nucleon cross 
section~\cite{Bertone:2007xj,Barger:2008qd,Belanger:2008gy,Cohen:2010gj}. In this 
paper we perform a model-independent classification of dark matter candidates whose 
interactions with nucleons are primarily spin-dependent. We study renormalizable 
theories where dark matter scattering is elastic and arises at tree-level. For each 
theory we calculate at lowest order the form of the effective operator that governs 
WIMP-nucleon scattering, and determine whether it leads to spin-dependent or 
spin-independent interactions. Our results are in excellent agreement with what 
would be expected from a naive operator analysis. We find that the dark matter 
candidates where spin-dependent interactions can naturally dominate are Majorana 
fermions and real vector bosons, so that the dark matter particle is its own 
anti-particle. However, the converse is not true. Specifically, the fact that the 
dark matter particle is a Majorana fermion or real vector boson is not by itself 
enough to guarantee that spin-dependent interactions dominate.

If the WIMP-nucleon cross section is primarily spin-dependent, scalar dark matter 
is disfavored. Dirac fermion and complex vector boson dark matter are also 
disfavored, except for very specific choices of quantum numbers. If dark matter 
is composed of scalars, WIMP-nucleon scattering is purely spin-independent. The 
interactions of Dirac fermions and complex vector bosons with nucleons always 
tend to have have a significant spin-independent component, unless the dark 
matter sector possesses a discrete symmetry similar in nature to charge 
conjugation, but under which the SM fields are invariant.
 
Theories where the tree-level WIMP-nucleon cross section is primarily 
spin-dependent share a feature which is potentially of great significance
for colliders. We find that any such theory must contain either
\begin{itemize}
  \item 
 new particles at the weak scale charged under at least one of 
 the SM gauge groups, or
  \item 
a $Z'$ gauge boson with mass at the TeV scale or below.
 \end{itemize} 
We show that in the region of parameter space that is of interest for 
current direct detection experiments, the masses of these new particles 
naturally lie in a range that is kinematically accessible to the Large 
Hadron Collider (LHC). The results of direct detection experiments that 
are looking for WIMPs with spin-dependent interactions with nuclei are 
therefore highly correlated with dark matter searches at the LHC.

The layout of the paper is as follows. In section 2, we distinguish using 
an effective field theory approach the dark matter candidates that have a 
primarily spin-dependent WIMP-nucleon cross section. We classify the 
various models in terms of the spin and quantum numbers of both the dark 
matter particle and of the intermediate particle mediating the 
interaction. We highlight the favored scenarios in section 3, and analyze 
in greater detail their consequences. In section 4, we investigate the 
link between direct detection experiments and collider experiments. We 
conclude in section 5.

\section{Classification of models}
\subsection{Operator Analysis}

In this section we identify and classify theories that have predominantly 
spin-dependent couplings to nucleons. WIMP dark matter is constrained to be 
neutral under both electromagnetism and color~\cite{Starkman:1990nj}. This 
implies that in renormalizable theories WIMPs do not scatter off gluons at 
tree-level. Our approach will therefore be to first identify all the 
renormalizable theories which generate tree-level WIMP-quark scattering. For each 
such theory we find the form of the effective operators that contribute to the 
WIMP-nucleon cross section. Each operator leads to either spin-dependent or 
spin-independent scattering, enabling us to distinguish theories where 
spin-dependent interactions dominate. We limit our analysis to purely elastic 
scattering, leaving the more complicated cases of inelastic dark matter 
\cite{TuckerSmith:2001hy,TuckerSmith:2004jv,Cui:2009xq} and form factor dark 
matter \cite{Feldstein:2009tr,Chang:2009yt,Bai:2009cd} for future work.
-
However, before proceeding, we first perform an operator analysis of
the various possibilities. Similar operator studies have been
performed in~\cite{Barger:2008qd,Beltran:2008xg}. The matrix element
for dark matter-nucleus scattering will in general involve the
expectation values of quark bilinear operators inside the nucleus of
the form $\langle N|\,\bar{q}\, \Gamma q\,|N\rangle$, where $\Gamma$
represents any of $\gamma^5$, $\gamma^\mu$, $\gamma^\mu \gamma^5$,
$\,\sigma^{\mu \nu}$ or simply the identity $\mathds{1}$. Of these,
$\bar{q} \gamma^5 q$ always leads to cross sections that are velocity
suppressed in the non-relativistic limit, and is therefore typically
sub-dominant. Furthermore, the operators $\bar{q} q$, $\bar{q}
\gamma^5 q$ and $\bar{q} \,\sigma^{\mu \nu} q$ violate the approximate
chiral symmetry of QCD, leading to cross sections that are suppressed
by the quark masses (unless there are additional sources of chiral
symmetry breaking in the theory). For these reasons the operators
$\bar{q} \gamma^\mu q$ and $\bar{q} \gamma^\mu \gamma^5 q$, when
present, in general tend to dominate the cross section. Of these, the
temporal component of $\bar{q} \gamma^\mu q$ leads to spin-independent
interactions, and the spatial component of $\bar{q} \gamma^\mu
\gamma^5 q$ leads to spin-dependent interactions.  The spatial
components of $\bar{q} \gamma^\mu q$ and the temporal component of
$\bar{q} \gamma^\mu \gamma^5 q$ are velocity suppressed.  The problem
of identifying theories where the dominant interactions of dark matter
with nucleons are sizeable and spin-dependent is therefore largely
equivalent to finding theories where the couplings of dark matter to
quarks are such that the operator $\bar{q} \gamma^\mu \gamma^5 q$ is
present in the low energy effective Lagrangian, whereas the operator
$\bar{q} \gamma^\mu q$ is not.

However, in some cases neither $\bar{q} \gamma^\mu q$ nor $\bar{q}
\gamma^\mu \gamma^5 q$ is generated at leading order. The dominant
contribution to the cross section may then arise from one or more of the
operators which are chirality suppressed. Of these $\bar{q} q$ always
generates spin-independent interactions while $\bar{q} \,\sigma^{\mu \nu} q$
always generates spin-dependent interactions in the non-relativistic
limit. The operator $\bar{q} \gamma^5 q$ generates spin-independent
interactions, but these are additionally velocity suppressed and usually
negligible.

Consider first the case of scalar dark matter. What are the possible
operators that involve scalars coupling to $\bar{q} \gamma^\mu
\gamma^5 q$ at leading order? Consider a complex scalar field, which we 
denote by $\phi$. One obvious candidate operator is
\begin{align}
  \phi^\dagger \partial_{\mu} \phi \; \bar{q} \gamma^\mu \gamma^5 q.
\end{align}
However, for both temporal and spatial $\mu$ the corresponding matrix
element is velocity suppressed.  It is straightforward to verify that
this result applies in general to any operator that couples scalar
dark matter to $\bar{q} \gamma^\mu \gamma^5 q$. No such constraint
arises in the cases of the operator $\bar{q} \gamma^\mu q$. For this
reason, the dominant interactions of complex scalar dark matter are in 
general spin-independent. 

The case of real scalar dark matter is slightly different. After 
integration by parts the operator above can be rewritten as
\begin{align}
  \phi \partial_{\mu} \phi \; \bar{q} \gamma^\mu q 
  \quad \rightarrow \quad
  \phi^2 \; \partial_{\mu} \left( \bar{q} \gamma^\mu q
  \right).
\end{align}
This operator does not contribute to the tree-level WIMP-nucleon cross 
section, as a consequence of the equation of motion (since $\bar{q} \gamma^\mu q$
corresponds to a conserved current). In such a scenario 
the leading contribution to the cross section is expected to arise from 
the chirality suppressed operator 
\begin{align}
    \phi^2 \bar{q} q.
\end{align}
Therefore, in the case of scalar dark matter, WIMP-nucleon scattering is 
in general spin-independent.

We move on to the case of fermionic dark matter. If we denote the dark
matter candidate by a four component spinor $\chi$,
the obvious candidate operators at leading order are
\begin{align}
  \bar{\chi} \gamma_{\mu} \chi \; \bar{q} \gamma^\mu \gamma^5 q
  \quad
  {\rm and}
  \quad
  \bar{\chi} \gamma_{\mu} \gamma^5 \chi \; \bar{q} \gamma^\mu \gamma^5
  q.
\end{align}
While the first of these operators is velocity suppressed, the
second does indeed lead to spin-dependent interactions.
However, for spin-dependent interactions to dominate, the operator
\begin{align}
  \bar{\chi} \gamma_{\mu} \chi \; \bar{q} \gamma^\mu q
\end{align}
must be absent. Now, if $\chi$ is a Majorana fermion, then $\bar{\chi} 
\gamma_{\mu} \chi$ identically vanishes and so this operator is automatically 
absent. However, in the case of Dirac fermions, in the absence of any symmetry to 
forbid this operator we expect that it will be present. One symmetry that can 
forbid this operator is a discrete symmetry under which the dark matter field 
$\chi$ is interchanged with $\chi^c$, its charge conjugate, while the SM fields 
are invariant. If the theory does not possess such a symmetry, we expect that the 
WIMP-nucleon scattering cross section will have a sizeable spin-independent 
component. As for the operator
\begin{align}
  \bar{\chi} \gamma_{\mu} \gamma^5 \chi \; \bar{q} \gamma^\mu q
\end{align}
which also gives rise to spin-independent interactions, it
turns out that the corresponding matrix elements are velocity
suppressed.

The final possibility is that the dark matter particle is a vector boson. We 
first consider the case where the corresponding vector field is real, so 
that the dark matter particle is its own anti-particle. If we denote the 
dark matter candidate by $B_{\mu}$, in the non-relativistic limit the 
physical degrees of freedom correspond to spatial $\mu$. The operator
\begin{align}
\epsilon_{\mu \nu \lambda \sigma} B^{\nu} \partial^{\lambda} B^{\sigma}
\;
\bar{q} \gamma^\mu \gamma^5 q
\end{align}
gives rise to spin-dependent interactions. The corresponding operator
but with $\bar{q} \gamma^\mu \gamma^5 q$ replaced by $\bar{q}
\gamma^\mu q$ is velocity suppressed. There also exist other operators that
arise at the same order such as
\begin{align}
  B^{\nu} \partial_{\mu} B_{\nu} \; \bar{q} \gamma^\mu \gamma^5 q 
  \quad
  {\rm and} \quad
B^{\nu} \partial_{\mu} B_{\nu} \; \bar{q} \gamma^\mu q.
\end{align}
However, these do not contribute significantly to the tree-level cross 
section. The effects of the first operator are velocity suppressed, while 
the second does not contribute significantly as a consequence of the 
equation of motion. 
Other operators that arise at the same order include
\begin{align}
B^{\nu} \partial_{\nu} B_{\mu} \; \bar{q} \gamma^\mu \gamma^5 q
\quad
{\rm and} \quad
B^{\nu} \partial_{\nu} B_{\mu} \; \bar{q} \gamma^\mu q.
\end{align}
However, it is straightforward to verify that these lead to interactions
that are always velocity suppressed. Therefore real vector boson dark
matter naturally leads to spin-dependent interactions.

Finally we come to the case of complex vector dark matter. The operator
\begin{align}
\epsilon^{\mu \nu \lambda \sigma} {B_{\nu}}^{\dagger} \partial_{\lambda} 
B_{\sigma} \; \bar{q} \gamma_\mu \gamma^5 q
\end{align}
gives rise to spin-dependent interactions. However, the operator
\begin{align}
B^{\nu} \partial_{\mu} {B_{\nu}}^{\dagger} \; \bar{q} \gamma^\mu q.
\end{align}
can now generate spin-independent interactions, and must be forbidden by a 
symmetry for spin-dependent interactions to dominate. One possibility is a 
discrete symmetry under which the vector field is interchanged with its charge 
conjugate $B_\mu \leftrightarrow - {B_{\mu}^\dagger}$, while the SM fields are 
invariant. In the absence of such a symmetry we expect that the WIMP-nucleon 
cross section will in general have a sizeable spin-independent component.

Our conclusions from the operator analysis are that scalar dark matter always 
leads to spin-independent interactions, while Majorana fermion and real vector 
boson dark matter naturally lead to spin-dependent interactions in the chiral 
limit. In the cases where dark matter is composed of Dirac fermions or complex 
vector bosons, we expect that the WIMP-nucleon cross section will in general have 
a sizeable spin-independent component, except for very specific choices of 
quantum numbers. In what follows we firm up this conclusion by studying each of 
these cases in more detail. Our approach will be to consider all renormalizable 
theories which lead to tree-level WIMP-quark scattering. For each theory we 
calculate the quark bilinears that appear in the effective operators which 
contribute to the WIMP-nucleon cross section, enabling us to immediately identify 
those theories where scattering is primarily spin-dependent.

\subsection{Scalar Dark Matter}

\begin{figure}[htp]
  \vspace{0.3in}
  \begin{center}
    \begin{fmffile}{outline-scalar}
      \subfloat[]{
      \label{fig:s1a}
      \begin{fmfgraph*}(30,20)
	\fmfleftn{ia}{2} \fmfrightn{oa}{2}
	\fmflabel{$\phi,\phi^\dagger$}{ia1}
	\fmflabel{$\phi,\phi^\dagger$}{ia2}
	\fmflabel{$q$}{oa1}
	\fmflabel{$q$}{oa2}
	\fmf{dashes}{ia1,va1,ia2}
	\fmf{boson,label=$Z$}{va1,va2}
	\fmf{plain}{oa1,va2,oa2}
      \end{fmfgraph*}}
      \hspace{0.5in}
      \subfloat[]{
      \label{fig:s1b}
      \begin{fmfgraph*}(30,20)
	\fmfleftn{ib}{2} \fmfrightn{ob}{2}
	\fmflabel{$\phi,\phi^\dagger$}{ib1}
	\fmflabel{$\phi,\phi^\dagger$}{ib2}
	\fmflabel{$q$}{ob1}
	\fmflabel{$q$}{ob2}
	\fmf{dashes}{ib1,vb1,ib2}
	\fmf{dashes, label=$h$}{vb1,vb2}
	\fmf{plain}{ob1,vb2,ob2}
      \end{fmfgraph*}}
      \\
      \vspace{0.2in}
      \subfloat[]{
      \label{fig:s1c}
      \begin{fmfgraph*}(30,20)
	\fmfleftn{ic}{2} \fmfrightn{oc}{2}
	\fmflabel{$\phi$}{ic1}
	\fmflabel{$\phi$}{ic2}
	\fmflabel{$q$}{oc1}
	\fmflabel{$q$}{oc2}
	\fmf{dashes}{ic1,vc1}
	\fmf{dashes}{ic2,vc2}
	\fmf{plain, label=$Q$}{vc1,vc2}
	\fmf{plain}{oc1,vc1}
	\fmf{plain}{vc2,oc2}
      \end{fmfgraph*}
      \hspace{0.5in}
      \begin{fmfgraph*}(30,20)
	\fmfleftn{id}{2} \fmfrightn{od}{2}
	\fmflabel{$\phi^\dagger$}{id1}
	\fmflabel{$\phi^\dagger$}{id2}
	\fmflabel{$q$}{od1}
	\fmflabel{$q$}{od2}
	\fmf{phantom}{id1,vd1}
	\fmf{phantom}{id2,vd2}
	\fmf{dashes,tension=0}{id2,vd1}
	\fmf{dashes,tension=0}{id1,vd2}
	\fmf{plain, label=$Q$, l.s=right}{vd1,vd2}
	\fmf{plain}{od1,vd1}
	\fmf{plain}{vd2,od2}
      \end{fmfgraph*}}
    \end{fmffile}
  \end{center}
  \caption{Scalar dark matter scattering diagrams}
  \label{fig:s1}
\end{figure}
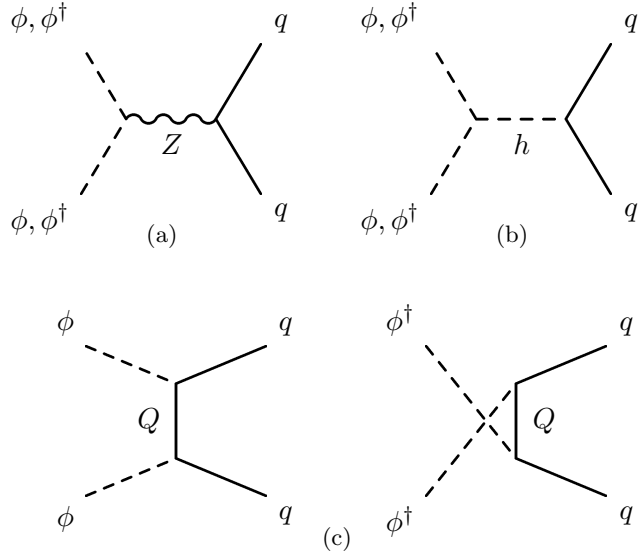

We begin by considering the case of scalar dark matter. In general the
scalar may be real, or it may be complex. Both cases are qualitatively
similar, but with subtle differences. In general, the tree-level
scattering of a scalar $\phi$ with quarks can occur through any of the
Feynman diagrams shown in Fig \ref{fig:s1}.

Our approach will be to start from the renormalizable Lagrangian that
corresponds to each diagram separately, integrate out the intermediate
particle, and find the form of the effective Lagrangian.  We are
particularly interested in interactions which generate a spin-dependent
($\bar{q}\gamma^\mu\gamma^5 q$) coupling to quarks.

\subsubsection*{Diagram \ref{fig:s1a}: t-channel vector exchange}

We note that this diagram only exists if the scalar particle is complex. Here the 
vector boson $Z$ being exchanged may correspond either to the $Z$ of the SM, or 
to a new $Z'$ which has been added to the SM. We do not consider the scenario 
where $Z$ corresponds to the SM photon since the photon always couples to the 
vector bilinear $\bar{q} \gamma^\mu q$, and we expect spin-independent scattering 
will dominate. At the renormalizable level, the form of the interaction between a 
complex scalar and a vector boson is fixed. In unitary gauge,
\begin{align}
  \mathcal{L}
  &=
  -\frac{1}{4}\mathcal{F}^{\mu\nu}\mathcal{F}_{\mu\nu}
  +\frac{1}{2}m_Z^2\;Z^{\mu}Z_{\mu}
  +a(\phi^{\dagger}\partial_{\mu}\phi-\phi\partial_{\mu}\phi^{\dagger})
  Z^{\mu}
  +\bar{q}\gamma^{\mu}(\alpha-\beta\gamma^5)q\;Z_{\mu}.
\end{align}
We now integrate out the vector particle using its equation of motion,
\begin{align}
  Z_{\mu} &=
  \left[
  (\partial^2+m_Z^2)g^{\beta\mu}
  -\partial^{\mu}\partial^{\beta}
  \right]^{-1}
  \left[
  -a(\phi^{\dagger}\partial^{\beta}\phi
  -\phi\partial^{\beta}\phi^{\dagger})
  -\bar{q}\gamma^{\beta}(\alpha-\beta\gamma^5)q
  \right].
\end{align}
The inverse of the differential operator above can be simplified using
\begin{align}
  \left[
  (\partial^2+m_Z^2)g^{\beta\mu}
  -\partial^{\mu}\partial^{\beta}
  \right]^{-1}
  &= \frac{g_{\beta\mu}+
  \frac{\partial_\beta\partial_\mu}{m_Z^2}}
  {\partial^2+m_Z^2}.
\end{align}
In order to elucidate the operator structure, we expand the above
in powers of $\partial/m_Z$, keeping only the zeroth and first order
terms. For t-channel processes such as this one, this corresponds to
assuming that the net momentum transfer in the
process is much less than $m_Z$, which is certainly valid in the
non-relativistic regime.
Under this approximation,
\begin{align}
  Z^{\mu}
  &\simeq \frac{1}{m_Z^2}
  \left[
  -a(\phi^{\dagger}\partial^{\mu}\phi
  -\phi\partial^{\mu}\phi^{\dagger})
  -\bar{q}\gamma^{\mu}(\alpha-\beta\gamma^5)q
  \right].
\end{align}
This leads to the effective Lagrangian
\begin{align}
  \mathcal{L}_{\rm eff}
  &\simeq
  \left[
  a(\phi^{\dagger}\partial^{\mu}\phi
  -\phi\partial^{\mu}\phi^{\dagger})
  +\bar{q}\gamma^{\mu}(\alpha-\beta\gamma^5)q
  \right]
  \frac{1}{2m_Z^2}
  \left[
  -a(\phi^{\dagger}\partial_{\mu}\phi
  -\phi\partial_{\mu}\phi^{\dagger})
  -\bar{q}\gamma_{\mu}(\alpha-\beta\gamma^5)q
  \right].
\end{align}
Keeping only terms in the effective Lagrangian relevant to our process,
this reduces to
\begin{align}
  \mathcal{L}_{\rm eff}
  &\simeq
  -\frac{a}{m_Z^2}
  (\phi^{\dagger}\partial_{\mu}\phi
  -\phi\partial_{\mu}\phi^{\dagger})\;
  \bar{q}\gamma^{\mu}(\alpha-\beta\gamma^5)q.
\end{align}
In the low energy limit the spatial components of the
derivative are suppressed. Therefore, the axial-vector contribution
vanishes in the non-relativistic limit and we are left with
\begin{align}
  \mathcal{L}_{\rm eff}
  &\simeq
  -\frac{a\alpha}{m_Z^2}
  (\phi^{\dagger}\partial_{\mu}\phi
  -\phi\partial_{\mu}\phi^{\dagger})\;
  \bar{q}\gamma^{\mu}q
  .
\end{align}
We see from this that, as expected, scattering is
spin-independent.

\subsubsection*{Diagram \ref{fig:s1b}: t-channel scalar exchange}

The dark matter field in this process may either be real or complex.
This constitutes the primary detection channel in several scalar dark
matter models
\cite{McDonald:1993ex,Burgess:2000yq,Barbieri:2005kf,LopezHonorez:2006gr,Majumdar:2006nt}.
If the dark matter field particle is complex, the general
renormalizable Lagrangian has the form
\begin{align}
  \mathcal{L}
  &=
  \frac12 (\partial h)^2 - \frac12 m_h^2 h^2
  -a\phi^{\dagger}\phi h
  - \bar{q}(\alpha-\beta\gamma^5)q h
  .
\end{align}
The case where the scalar field is real is very similar and may be
recovered simply by setting $\phi^\dagger = \phi$ in the above
Lagrangian. Here $h$ could represent the SM Higgs, or more
generally any real scalar that couples to quarks.
We follow the same procedure as earlier. Using the
equation of motion for the intermediate scalar $h$, we integrate
it out,
\begin{align}
  h&= (\partial^2+m_h^2)^{-1}
  \left[
  -a\phi^{\dagger}\phi-\bar{q}(\alpha-\beta\gamma^5)q
  \right]
  .
\end{align}
As before, working in an expansion in powers of $\partial/m_h$ and
neglecting terms of order $\partial^2/m_h^2$ or higher, we have
\begin{align}
  h &= -\frac{a\phi^{\dagger}\phi
  +\bar{q}(\alpha-\beta\gamma^5)q}{m_h^2}
  .
\end{align}
The relevant part of the effective Lagrangian is then
\begin{align}
  \mathcal{L}_{\rm eff}
  &\simeq
  \frac{\phi^{\dagger}\phi}{m_h^2}
  \left[
   a\alpha\,
  \bar{q}q
  -
  a \beta\,
  \bar{q}\gamma^5q
  \right]
  .
\end{align}
As noted earlier, the spinor
$\bar{q}\gamma^5 q$ vanishes in the non-relativistic limit. We are
left with
\begin{align}
  \mathcal{L}_{\rm eff}
  &\simeq
  \frac{a\alpha}{m_h^2}
  \phi^{\dagger}\phi\;
  \bar{q}\,q
  .
\end{align}
which generates purely spin-independent interactions.
The reason for the apparent disagreement with the naive operator
analysis of the previous subsection is that the interaction
Lagrangian we started from here explicitly breaks
chiral symmetry. Therefore
it is not surprising that the leading
operator in the effective theory also breaks chiral symmetry.

\subsubsection*{Diagram \ref{fig:s1c}: s- and u-channel fermion exchange}

In this case there are two distinct possibilities, depending on whether 
the scalar field is real or complex. For a complex scalar field there is 
only one diagram, which may either be s-channel or u-channel, depending on 
whether it is the particle or anti-particle being scattered. On the other 
hand, for a real scalar both s- and u-channel diagrams are present.

For the complex scalar, the calculation proceeds as follows.  We
start with the general form of the Lagrangian,
\begin{align}
  \mathcal{L}
  &=
  \bar{Q}(i\slashed{\partial}-m_Q)Q
  -\bar{q}(\alpha-\beta\gamma^5)Q\phi^{\dagger}
  -\bar{Q}(\alpha^*+\beta^*\gamma^5)q\phi
  .
\end{align}
The equation of motion for $Q$ takes the form,
\begin{align}
  Q
  &=\frac{i\slashed{\partial}+m_Q}{\partial^2+m_Q^2}
  \left[
  -(\alpha^*+\beta^*\gamma^5)q\phi
  \right]
  .
\end{align}
Derivatives acting on the quark fields can be neglected, since these 
effects are generally suppressed by powers of $\Lambda_{\rm QCD}/m_Q$ (or 
$\Lambda_{\rm QCD}/m_{\phi}$) relative to other contributions.  We expand 
in $\partial/m_Q$ as before, keeping only zeroth order and first order 
terms. For s-channel and u-channel processes in the non-relativistic 
limit, this corresponds to the assumption that the WIMP mass squared is 
much less than the mass squared of the particle being exchanged, 
$m_{\phi}^2 \ll m_{Q}^2$. Although $m_{\phi} < m_{Q}$ is required for WIMP 
decays to be kinematically forbidden, it does not follow that $m_{\phi}^2 
\ll m_{Q}^2$ is necessarily satisfied. However, this approximation 
suffices to determine the leading term in the low-energy effective 
interaction that emerges from this class of theories, and determine 
whether it leads to spin-dependent or spin-independent interactions. We 
will relax this assumption in the next section. Then
\begin{align}
  Q
  &=
  -m_Q(\alpha^*+\beta^*\gamma^5)q
  \frac{1}{m_Q^2}\phi
  -
  \gamma^\mu(\alpha^*+\beta^*\gamma^5)q
  \frac{i\partial_\mu}{m_Q^2}\phi
  .
\end{align}
The relevant terms in the effective Lagrangian are given by
\begin{align}
  \mathcal{L}_{\rm eff}
  &\simeq
  \frac{1}{m_Q^2}
  \bar{q}(\alpha-\beta\gamma^5)\phi^{\dagger}
  \left(
  m_Q(\alpha^*+\beta^*\gamma^5)q
  \phi
  +
  i\gamma^\mu(\alpha^*+\beta^*\gamma^5)q
  \partial_\mu\phi
  \right)
  .
\end{align}
In the non-relativistic limit the derivative picks out the time
direction, and therefore only the vector quark current survives in the
second term above. We are left with
\begin{align}
  \mathcal{L}_{\rm eff}
  &\simeq
  \frac{1}{m_Q}
  (|\alpha|^2-|\beta|^2)\;
  \bar{q}q \;
  \phi^{\dagger}
  \phi
  +
  \frac{i}{m_Q^2}
  (|\alpha|^2+|\beta|^2)
  \;\bar{q}\gamma^\mu q \;
  \phi^{\dagger}
  \partial_\mu\phi
  .
\end{align}
We see from this that the only interactions that are generated have the 
scalar and vector forms, both of which lead to spin-independent
scattering. As expected, in the chiral limit (when $\alpha= 
\pm \beta$) the scalar contribution vanishes and we are left with just 
the vector interaction. For the real scalar, 
the effective Lagrangian is obtained by simply setting $\phi^\dagger = 
\phi$ in the equation above. In this case the second term in the equation 
above does not contribute, as may be verified by integrating by parts. 
Then the leading contribution to the cross section 
arises from a chiral symmetry breaking effect, as expected from our 
operator analysis.
In summary,  we never
generate sizeable spin-dependent interactions at low energies in the case
of scalar dark matter.

\subsection{Fermionic Dark Matter}

We now move on to the case of fermionic dark matter. It is convenient to
consider the cases of the Dirac fermion and Majorana fermion separately.

\subsubsection{Dirac Fermion Dark Matter}

\begin{figure}[htp]
  \begin{center}
    \vspace{0.3in}
    \begin{fmffile}{outline-fermion}
      \subfloat[]{
      \label{fig:f1a}
      \begin{fmfgraph*}(30,20)
	\fmfleftn{ia}{2} \fmfrightn{oa}{2}
	\fmflabel{$\chi,\chi^c$}{ia1}
	\fmflabel{$\chi,\chi^c$}{ia2}
	\fmflabel{$q$}{oa1}
	\fmflabel{$q$}{oa2}
	\fmf{plain}{ia1,va1,ia2}
	\fmf{boson,label=$Z$}{va1,va2}
	\fmf{plain}{oa1,va2,oa2}
      \end{fmfgraph*}}
      \hspace{0.4in}
      \subfloat[]{
      \label{fig:f1b}
      \begin{fmfgraph*}(30,20)
	\fmfleftn{ib}{2} \fmfrightn{ob}{2}
	\fmflabel{$\chi,\chi^c$}{ib1}
	\fmflabel{$\chi,\chi^c$}{ib2}
	\fmflabel{$q$}{ob1}
	\fmflabel{$q$}{ob2}
	\fmf{plain}{ib1,vb1,ib2}
	\fmf{dashes, label=$h$}{vb1,vb2}
	\fmf{plain}{ob1,vb2,ob2}
      \end{fmfgraph*}}
      \vspace{0.2in}
      \\
      \subfloat[]{
      \label{fig:f1c}
      \begin{fmfgraph*}(30,20)
	\fmfleftn{ic}{2} \fmfrightn{oc}{2}
	\fmflabel{$\chi$}{ic1}
	\fmflabel{$\chi$}{ic2}
	\fmflabel{$q$}{oc1}
	\fmflabel{$q$}{oc2}
	\fmf{plain}{ic1,vc1}
	\fmf{plain}{vc2,ic2}
	\fmf{boson, label=$X$}{vc1,vc2}
	\fmf{phantom}{oc1,vc1}
	\fmf{phantom}{vc2,oc2}
	\fmf{plain,tension=0}{oc1,vc2}
	\fmf{plain,tension=0}{vc1,oc2}
      \end{fmfgraph*}
      \hspace{0.4in}
      \label{fig:f1ca}
      \begin{fmfgraph*}(30,20)
	\fmfleftn{ic}{2} \fmfrightn{oc}{2}
	\fmflabel{$\chi^c$}{ic1}
	\fmflabel{$\chi^c$}{ic2}
	\fmflabel{$q$}{oc1}
	\fmflabel{$q$}{oc2}
	\fmf{plain}{vc1,ic1}
	\fmf{plain}{ic2,vc2}
	\fmf{boson, label=$X$}{vc1,vc2}
	\fmf{plain}{oc1,vc1}
	\fmf{plain}{vc2,oc2}
      \end{fmfgraph*}}
      \\
      \vspace{0.2in}
      \subfloat[]{
      \label{fig:f1d}
      \begin{fmfgraph*}(30,20)
	\fmfleftn{id}{2} \fmfrightn{od}{2}
	\fmflabel{$\chi$}{id1}
	\fmflabel{$\chi$}{id2}
	\fmflabel{$q$}{od1}
	\fmflabel{$q$}{od2}
	\fmf{plain}{id1,vd1}
	\fmf{plain}{vd2,id2}
	\fmf{dashes, label=$\Phi$}{vd1,vd2}
	\fmf{phantom}{od1,vd1}
	\fmf{phantom}{vd2,od2}
	\fmf{plain,tension=0}{od1,vd2}
	\fmf{plain,tension=0}{vd1,od2}
      \end{fmfgraph*}
      \hspace{0.4in}
      \label{fig:f1da}
      \begin{fmfgraph*}(30,20)
	\fmfleftn{id}{2} \fmfrightn{od}{2}
	\fmflabel{$\chi^c$}{id1}
	\fmflabel{$\chi^c$}{id2}
	\fmflabel{$q$}{od1}
	\fmflabel{$q$}{od2}
	\fmf{plain}{vd1,id1}
	\fmf{plain}{id2,vd2}
	\fmf{dashes, label=$\Phi$}{vd1,vd2}
	\fmf{plain}{od1,vd1}
	\fmf{plain}{vd2,od2}
      \end{fmfgraph*}}
    \end{fmffile}
  \end{center}
  \caption{Dirac fermion dark matter scattering diagrams}
\end{figure}
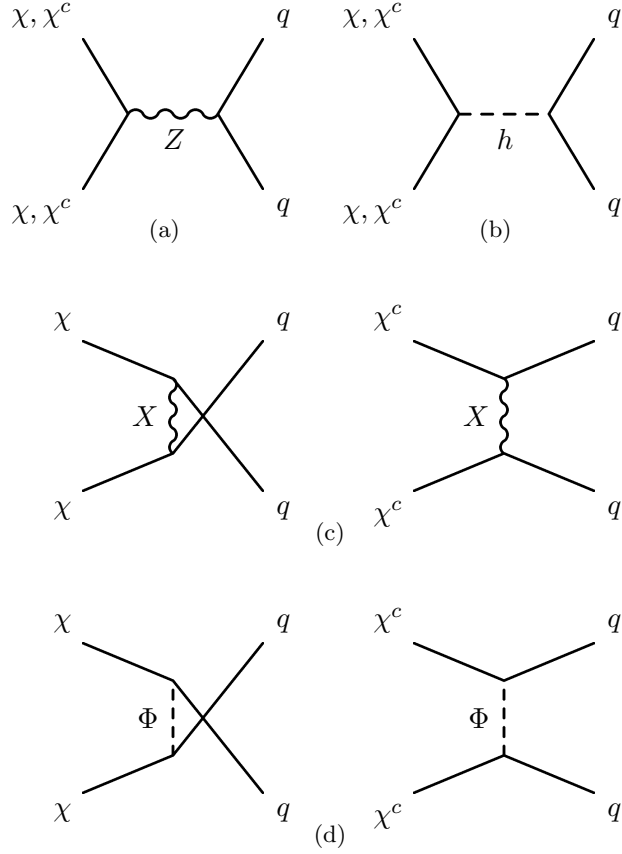

We begin with the case where dark matter consists of Dirac fermions. In what 
follows we require that the Lagrangian has a symmetry under which the Dirac field 
$\chi$ can be arbitrarily rephased, $\chi \rightarrow \chi \; {\rm exp} \left( i 
\theta \right)$. This is necessary to ensure that the Dirac nature of $\chi$ is 
maintained at loop level, so that $\chi$ does not split into two Majorana states. 
At tree-level the interaction of the dark matter particle with quarks can arise 
in any of four different ways as shown in the diagrams below. Diagrams 
\ref{fig:f1a} and \ref{fig:f1b} arise in some specific theories with Dirac 
fermion dark matter~\cite{Agashe:2004bm,Brahm:1989jh}.

\subsubsection* {Diagram \ref{fig:f1a}: t-channel vector exchange}
The most general renormalizable Lagrangian corresponding to this
process takes the form
\begin{align}
  \mathcal{L}&=
  -\frac{1}{4}\mathcal{F}^{\mu\nu}\mathcal{F}_{\mu\nu}
  +\frac{1}{2}m_Z^2\;Z^{\mu}Z_{\mu}
  +\bar{\chi}\gamma^{\mu}
  (\alpha-\beta\gamma^5)\chi Z_{\mu}
  +\bar{q}\gamma^{\mu}(\widetilde\alpha-\widetilde\beta\gamma^5)q Z_{\mu}
  .
\end{align}
The analysis is similar to that of diagram \ref{fig:s1a}.  In particular,
since the external states are still non-relativistic, the analysis of
integrating out the mediator particle is exactly same as before. Once
again we neglect terms of order $\partial^2/m_Z^2$, leading to
\begin{align}
  Z_{\mu}
  &\simeq
  \frac{g_{\mu\nu}}{m_Z^2}
  \left[
  -\bar{\chi}\gamma^{\nu}(\alpha-\beta\gamma^5)\chi
  -\bar{q}\gamma^{\nu}(\widetilde\alpha-\widetilde\beta\gamma^5)q
  \right]
  .
\end{align}
The effective Lagrangian can be immediately read off as
\begin{align}
  \mathcal{L}_{\rm eff}
  &=\left[
  \bar{\chi}\gamma^{\mu}(\alpha-\beta\gamma^5)\chi
  +\bar{q}\gamma^{\mu}(\widetilde\alpha-\widetilde\beta\gamma^5)q
  \right]
  \frac{g_{\mu\nu}}{2m_Z^2}
  \left[
  -\bar{\chi}\gamma^{\nu}(\alpha-\beta\gamma^5)\chi
  -\bar{q}\gamma^{\nu}(\widetilde\alpha-\widetilde\beta\gamma^5)q
  \right]
  \label{eq:leffdfa}
  .
\end{align}
We note that the mixed terms above do not survive in the
non-relativistic limit.
Keeping only terms relevant to our process, we are left with
\begin{align}
  \mathcal{L}_{\rm eff}
  &\simeq -\frac{1}{m_Z^2}
  \left[
  \alpha\widetilde\alpha\;
  \bar{\chi}\gamma^{\mu}\chi \;
  \bar{q}\gamma_{\mu}q
  +\beta\widetilde\beta \;
  \bar{\chi}\gamma^{\mu}\gamma^5\chi \;
  \bar{q}\gamma_{\mu}\gamma^5q
  \right]
  .
\end{align}
We see that in general both spin-dependent and spin-independent interaction terms 
arise. However, if either $\alpha$ or $\widetilde{\alpha}$ is zero, the 
interaction is purely spin-dependent. The case $\alpha = 0$ corresponds to the 
limit where the dark matter sector possesses a symmetry under which $\chi 
\leftrightarrow \chi^c$, while the SM fields are invariant. We will consider this 
possibility in greater detail in the next section.

\subsubsection* {Diagram \ref{fig:f1b}: t-channel scalar exchange}
The relevant part of the Lagrangian takes the form
\begin{align}
  \mathcal{L}
  &= 
  \frac12 (\partial h)^2 - \frac12 m_h^2 h^2
  -\bar{\chi}(\alpha-\beta\gamma^5)\chi h
  -\bar{q}(\widetilde\alpha-  \widetilde\beta\gamma^5)q h
  .
\end{align}
We follow the same procedure as earlier to integrate out $h$. We solve
the equation of motion for $h$, and ignore terms of order
$\partial^2/m_h^2$ and higher,
\begin{align}
  h &\simeq -\frac{1}{m_h^2}
  \left[
  \bar{\chi}(\alpha-\beta\gamma^5)\chi
  +\bar{q}(\wt\alpha-\wt\beta\gamma^5)q
  \right]
  .
\end{align}
This leads to the following effective Lagrangian
\begin{align}
  \mathcal{L}_{\rm eff}
  &=
  \left[
  \bar{\chi}(\alpha-\beta\gamma^5)\chi
  +\bar{q}(\widetilde\alpha-\widetilde\beta\gamma^5)q
  \right]
  \frac{1}{2m_h^2}
  \left[
  \bar{\chi}(\alpha-\beta\gamma^5)\chi
  +\bar{q}(\widetilde\alpha-\widetilde\beta\gamma^5)q
  \right]
  .
\end{align}
Keeping only the terms relevant to our process that survive in the
non-relativistic limit,
\begin{align}
\mathcal{L}_{\rm eff}  &\simeq
  \frac{\alpha\wt\alpha}{m_h^2}
  \bar{\chi}\chi\;\bar{q}q
  .
\end{align}
This leads to purely spin-independent interactions that are suppressed in
the limit of exact chiral symmetry. As before, the reason for the apparent
disagreement with the naive operator analysis of the previous subsection
is that the interaction Lagrangian we started from explicitly breaks
chiral symmetry.

\subsubsection* {Diagram \ref{fig:f1c}: s- and u-channel vector exchange}
We start from the following Lagrangian,
\begin{align}
  \mathcal{L}
  &=
  -\frac{1}{2}\left|\partial_\mu X_\nu - \partial_\nu X_\mu\right|^2
  +m_X^2\;X^\dagger_\mu X^\mu
  +\bar{\chi}\gamma^{\mu}(\alpha-\beta\gamma^5)q
  X_{\mu}
  +\bar{q}\gamma^{\mu}(\alpha^*-\beta^*\gamma^5)\chi
  X_{\mu}^\dagger
  .
\end{align}
We can integrate out the colored vector boson $X_\mu$ using its
equation of motion. Dropping terms of order $\partial^2/m_X^2$ and
restricting to terms relevant to our process, we get
\begin{align}
  \mathcal{L}_{\rm eff}
  &\simeq -\frac{1}{m_X^2}
  \left[
  \bar{\chi}\gamma^{\nu}(\alpha-\beta\gamma^5)q\;
  \bar{q}\gamma_{\nu}(\alpha^*-\beta^*\gamma^5)\chi
  \right]
  .
\end{align}
We can use the Fierz identities listed in Appendix \ref{sec:fierz} to
rewrite the bilinears above.  After dropping terms which vanish in the
non-relativistic limit, the effective Lagrangian takes the form
\begin{align}
  \mathcal{L}
  &\simeq
  \frac{1}{m_X^2}
  \left[
  (|\alpha|^2-|\beta|^2)\bar{q}q\bar{\chi}\chi
  -\frac12(|\alpha|^2+|\beta|^2)
  (\bar{q}\gamma^{\mu}q
  \bar{\chi}\gamma_{\mu}\chi
  +\bar{q}\gamma^{\mu}\gamma^{5}q
  \bar{\chi}\gamma_{\mu}\gamma^{5}\chi)
  \right]
  \label{eq:diracX}
  .
\end{align}
We see that in general both spin-dependent and spin-independent interactions are 
generated, and with comparable magnitudes. The scalar term vanishes in the chiral 
limit ($\alpha=\pm \beta$). This conclusion is not unexpected because there are 
no values of the parameters $\alpha$ and $\beta$ for which the theory enjoys a 
$\chi \leftrightarrow \chi^c$ symmetry. While it is in fact possible to 
incorporate such a symmetry in a manner consistent with the rephasing symmetry of 
$\chi$, this requires adding additional fields to the theory. We will therefore 
not consider this possibility further.

\subsubsection* {Diagram \ref{fig:f1d}: s- and u-channel scalar exchange}
The part of the Lagrangian relevant to this process has the general
form
\begin{align}
  \mathcal{L}
  &=
  |\partial \Phi|^2-m_\Phi^2|\Phi|^2
  -\bar{\chi}(\alpha-\beta\gamma^5)q \Phi
  -\bar{q}(\alpha^*+\beta^*\gamma^5)\chi \Phi^{\dagger}
  .
\end{align}
Using the equation of motion for $\Phi$ (ignoring terms of
order $\partial^2/m_{\Phi}^2$), we get
\begin{align}
  \Phi
  \simeq -\frac{\bar{q}
  (\alpha^*+\beta^*\gamma^5)\chi}{m_\Phi^2}
  .
\end{align}
This leads to the effective Lagrangian
\begin{align}
  \mathcal{L}_{\rm eff}
  &\simeq
  \frac{1}{m_\Phi^2}
  \left[
  |\alpha|^2(\bar{\chi}q) (\bar{q}\chi)
  -|\beta|^2(\bar{\chi}\gamma^5q)
  (\bar{q}\gamma^5\chi)
  +\alpha\beta^*(\bar{\chi}q)(\bar{q}\gamma^5\chi)
  -\alpha^*\beta(\bar{\chi}\gamma^5q)(\bar{q}\chi)
  \right]
  .
\end{align}
After a Fierz rearrangement, neglecting terms which are velocity
suppressed we are left with
\begin{align}
  \mathcal{L}_\text{eff}
  &\simeq
  -\frac{1}{4m_\Phi^2}
  \left[
  (|\alpha|^2-|\beta|^2)(\bar{q}q\bar{\chi}\chi
  +\frac{1}{2}\bar{q}\,\sigma^{\mu\nu}q
  \bar{\chi}\,\sigma_{\mu\nu}\chi)
  +(|\alpha|^2+|\beta|^2)(\bar{q}\gamma^{\mu}q\bar{\chi}\gamma_{\mu}\chi
  -\bar{q}\gamma^{\mu}\gamma^{5}q
  \bar{\chi}\gamma_{\mu}\gamma^{5}\chi)\right]
  .
\end{align}
As in the previous case, we see that both spin-independent and spin-dependent 
interactions are necessarily generated. As expected, the scalar and tensor 
contributions vanish in the chiral limit. Once again, this conclusion could have 
been anticipated because there are no values of the parameters $\alpha$ and 
$\beta$ that are consistent with a $\chi \leftrightarrow \chi^c$ interchange 
symmetry. Incorporating such a symmetry without violating the rephasing symmetry 
of $\chi$ necessarily requires expanding upon this minimal field content, and we 
will therefore not expand on this possibility.

We see that in the case that the dark matter particle is a Dirac fermion, 
in the chiral limit most interactions necessarily generate both 
spin-dependent and spin-independent scattering. The solitary exception to 
this general rule arises in the case of vector boson exchange in the 
t-channel, and then only for very specific choices of charges. We will 
return to this possibility in the next section.

\subsubsection{Majorana Fermion Dark Matter}
\begin{figure}[htp]
  \begin{center}
    \begin{fmffile}{outline-majorana}
      \subfloat[]{
      \label{fig:fm1a}
      \begin{fmfgraph*}(30,20)
	\fmfleftn{ia}{2} \fmfrightn{oa}{2}
	\fmflabel{$\chi$}{ia1}
	\fmflabel{$\chi$}{ia2}
	\fmflabel{$q$}{oa1}
	\fmflabel{$q$}{oa2}
	\fmf{plain}{ia1,va1,ia2}
	\fmf{boson,label=$Z$}{va1,va2}
	\fmf{plain}{oa1,va2,oa2}
      \end{fmfgraph*}}
      \hspace{0.3in}
      \subfloat[]{
      \label{fig:fm1b}
      \begin{fmfgraph*}(30,20)
	\fmfleftn{ib}{2} \fmfrightn{ob}{2}
	\fmflabel{$\chi$}{ib1}
	\fmflabel{$\chi$}{ib2}
	\fmflabel{$q$}{ob1}
	\fmflabel{$q$}{ob2}
	\fmf{plain}{ib1,vb1,ib2}
	\fmf{dashes, label=$h$}{vb1,vb2}
	\fmf{plain}{ob1,vb2,ob2}
      \end{fmfgraph*}}
      \\
      \hspace{0.3in}
      \subfloat[]{
      \label{fig:fm1c}
      \begin{fmfgraph*}(30,20)
	\fmfleftn{ic}{2} \fmfrightn{oc}{2}
	\fmflabel{$\chi$}{ic1}
	\fmflabel{$\chi$}{ic2}
	\fmflabel{$q$}{oc1}
	\fmflabel{$q$}{oc2}
	\fmf{plain}{ic1,vc1}
	\fmf{plain}{vc2,ic2}
	\fmf{boson, label=$X$}{vc1,vc2}
	\fmf{phantom}{oc1,vc1}
	\fmf{phantom}{vc2,oc2}
	\fmf{plain,tension=0}{oc1,vc2}
	\fmf{plain,tension=0}{vc1,oc2}
      \end{fmfgraph*}
      \hspace{0.3in}
      \label{fig:fm1ca}
      \begin{fmfgraph*}(30,20)
	\fmfleftn{ic}{2} \fmfrightn{oc}{2}
	\fmflabel{$\chi$}{ic1}
	\fmflabel{$\chi$}{ic2}
	\fmflabel{$q$}{oc1}
	\fmflabel{$q$}{oc2}
	\fmf{plain}{vc1,ic1}
	\fmf{plain}{ic2,vc2}
	\fmf{boson, label=$X$}{vc1,vc2}
	\fmf{plain}{oc1,vc1}
	\fmf{plain}{vc2,oc2}
      \end{fmfgraph*}}
      \\
      \hspace{0.3in}
      \subfloat[]{
      \label{fig:fm1d}
      \begin{fmfgraph*}(30,20)
	\fmfleftn{id}{2} \fmfrightn{od}{2}
	\fmflabel{$\chi$}{id1}
	\fmflabel{$\chi$}{id2}
	\fmflabel{$q$}{od1}
	\fmflabel{$q$}{od2}
	\fmf{plain}{id1,vd1}
	\fmf{plain}{vd2,id2}
	\fmf{dashes, label=$\Phi$}{vd1,vd2}
	\fmf{phantom}{od1,vd1}
	\fmf{phantom}{vd2,od2}
	\fmf{plain,tension=0}{od1,vd2}
	\fmf{plain,tension=0}{vd1,od2}
      \end{fmfgraph*}
      \hspace{0.3in}
      \label{fig:fm1da}
      \begin{fmfgraph*}(30,20)
	\fmfleftn{id}{2} \fmfrightn{od}{2}
	\fmflabel{$\chi$}{id1}
	\fmflabel{$\chi$}{id2}
	\fmflabel{$q$}{od1}
	\fmflabel{$q$}{od2}
	\fmf{plain}{vd1,id1}
	\fmf{plain}{id2,vd2}
	\fmf{dashes, label=$\Phi$}{vd1,vd2}
	\fmf{plain}{od1,vd1}
	\fmf{plain}{vd2,od2}
      \end{fmfgraph*}}
    \end{fmffile}
  \end{center}
  \caption{Majorana fermion dark matter scattering diagrams}
\end{figure}
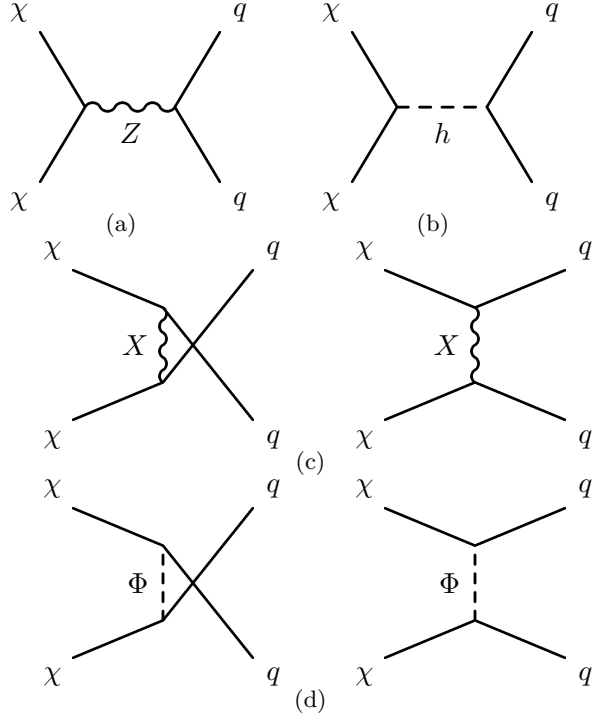
We now move on to the case where the dark matter is Majorana.
Due to the fact that the Majorana fermion is its own anti-particle,
those bilinears which are odd under charge conjugation vanish for
Majorana spinors.
\begin{align}
  \begin{array}{|c|c|}
    \hline
    \text{Bilinear} & \text{C}\\
    \hline
    \bar{\chi} \chi & +\\
    i\bar{\chi}\gamma^5 \chi&+\\
    \bar{\chi} \gamma^\mu\chi &- \\
    \bar{\chi} \gamma^\mu\gamma^5\chi &+\\
    \bar{\chi} \,\sigma^{\mu\nu}\chi & -\\
    \hline
  \end{array}
\end{align}
Therefore, for a Majorana fermion, the vector and the tensor bilinears
vanish. With this in mind, we can now use our earlier results from the
Dirac fermion case with only minor modifications.
Majorana fermions arise naturally in supersymmetric theories of dark
matter, in particular the processes corresponding to  diagrams
\ref{fig:fm1a}, \ref{fig:fm1b} and \ref{fig:fm1d}~\cite{Ellis:1983ew}.
\subsubsection* {Diagram \ref{fig:fm1a}: t-channel vector exchange}
Setting the vector coupling of
$\chi$ to zero in the effective Lagrangian for the Dirac case in
equation (\ref{eq:leffdfa}), we find
\begin{align}
  \mathcal{L}_{\rm eff}
  &=\left[
  \bar{\chi}\gamma^{\mu}(-\beta\gamma^5)\chi
  +\bar{q}\gamma^{\mu}(\widetilde\alpha-\widetilde\beta\gamma^5)q
  \right]
  \frac{g_{\mu\nu}}{2m_Z^2}
  \left[-
  \bar{\chi}\gamma^{\nu}(-\beta\gamma^5)\chi
  -\bar{q}\gamma^{\nu}(\widetilde\alpha-\widetilde\beta\gamma^5)q
  \right]
  .
\end{align}
Note that we have chosen a convention where assigning the canonically
normalized Majorana field unit charge under $Z$ corresponds to setting
$\beta=\frac12$.
Keeping only terms relevant to our process, and dropping interactions
which are velocity suppressed we see that
\begin{align}
  \mathcal{L}_{\rm eff}
  &\simeq -\frac{\beta\widetilde\beta}{m_Z^2}
  \bar{\chi}\gamma^{\mu}\gamma^5\chi
  \bar{q}\gamma_{\mu}\gamma^5q
  \label{eq:ma}
  .
\end{align}
We see that only spin-dependent
couplings are generated, in agreement with our operator analysis.
\subsubsection* {Diagram \ref{fig:fm1b}: t-channel scalar exchange}
The analysis here is very similar to that of the Dirac case considered
earlier, and also leads to interactions that
are purely spin-independent.
\subsubsection* {Diagram \ref{fig:fm1c}: s- and u-channel vector
exchange}
If we again work to zeroth order in $\partial^2/m_X^2$, we can just
copy the result from the Dirac case in equation (\ref{eq:diracX}),
dropping the vector interaction term. Hence
\begin{align}
  \mathcal{L}_{\rm eff}
  &\simeq
  \frac{1}{m_X^2}
  \left[
  (|\alpha|^2-|\beta|^2)\bar{q}q\bar{\chi}\chi
  -\frac12(|\alpha|^2+|\beta|^2)
  (\bar{q}\gamma^{\mu}\gamma^{5}q
  \bar{\chi}\gamma_{\mu}\gamma^{5}\chi)
  \right]
  .
\end{align}
The dominant interaction is spin-dependent. There is also a
spin-independent contribution which vanishes in the limit of exact chiral
symmetry ($\alpha=\pm\beta$), and which can therefore naturally be small.
Thus, this is another class of interactions which generates primarily
spin-dependent interactions.

\subsubsection* {Diagram \ref{fig:fm1d}: s- and u-channel scalar
exchange}
This case is again very similar to that of Dirac fermions
except for the vector and tensor couplings, which vanish for a
Majorana fermion. The effective Lagrangian is then
\begin{align}
  \mathcal{L}_{\rm eff}
  &\simeq
  -\frac{1}{4m_\Phi^2}
  \left[
  (|\alpha|^2-|\beta|^2)
  \bar{q}q\bar{\chi}\chi
  -(|\alpha|^2+|\beta|^2)
  \bar{q}\gamma^{\mu}\gamma^{5}q
  \bar{\chi}\gamma_{\mu}\gamma^{5}\chi
  \right]
  .
\end{align}
As expected from our operator analysis, this interaction
generates exclusively spin-dependent cross sections if chiral
symmetry ($\alpha=\pm\beta$) is imposed.

To summarize, when the dark matter particle is a Majorana fermion, the
non-relativistic WIMP-nucleon cross section is dominated by spin-dependent
interactions in the chiral limit, in perfect agreement with the operator
analysis.

\subsection{Vector Dark Matter}

We now consider the case of dark matter being a vector field, which may be real 
or complex. In the case that the vector field $B_\mu$ is complex, we require that 
the Lagrangian be invariant under the rephasing of $B_{\mu}$ by an arbitrary 
amount, $B_{\mu} \rightarrow B_{\mu} \; {\rm exp}\left(i \theta \right)$. This is 
to ensure that the degeneracy of the two real vector fields that constitute 
$B_\mu$ is radiatively stable.

\begin{figure}[htp]
  \begin{center}
    \vspace{0.3in}
    \begin{fmffile}{outline-vector}
      \subfloat[]{
      \label{fig:v1a}
      \begin{fmfgraph*}(30,20)
	\fmfleftn{ia}{2} \fmfrightn{oa}{2}
	\fmflabel{$B_\mu,B_\mu^\dagger$}{ia1}
	\fmflabel{$B_\nu,B_\nu^\dagger$}{ia2}
	\fmflabel{$q$}{oa1}
	\fmflabel{$q$}{oa2}
	\fmf{boson}{ia1,va1,ia2}
	\fmf{boson,label=$Z$}{va1,va2}
	\fmf{plain}{oa1,va2,oa2}
      \end{fmfgraph*}}
      \hspace{0.6in}
      \subfloat[]{
      \label{fig:v1b}
      \begin{fmfgraph*}(30,20)
	\fmfleftn{ib}{2} \fmfrightn{ob}{2}
	\fmflabel{$B_\mu,B_\mu^\dagger$}{ib1}
	\fmflabel{$B_\nu,B_\nu^\dagger$}{ib2}
	\fmflabel{$q$}{ob1}
	\fmflabel{$q$}{ob2}
	\fmf{boson}{ib1,vb1,ib2}
	\fmf{dashes, label=$h$}{vb1,vb2}
	\fmf{plain}{ob1,vb2,ob2}
      \end{fmfgraph*}}
      \\
      \subfloat[]{
      \label{fig:v1c}
      \begin{fmfgraph*}(30,20)
	\fmfleftn{ic}{2} \fmfrightn{oc}{2}
	\fmflabel{$B_\mu$}{ic1}
	\fmflabel{$B_\nu$}{ic2}
	\fmflabel{$q$}{oc1}
	\fmflabel{$q$}{oc2}
	\fmf{boson}{ic1,vc1}
	\fmf{boson}{ic2,vc2}
	\fmf{plain,label=$Q$}{vc1,vc2}
	\fmf{plain}{oc1,vc1}
	\fmf{plain}{vc2,oc2}
      \end{fmfgraph*}
      \hspace{0.6in}
      \begin{fmfgraph*}(30,20)
	\fmfleftn{id}{2} \fmfrightn{od}{2}
	\fmflabel{$B_\mu^\dagger$}{id1}
	\fmflabel{$B_\nu^\dagger$}{id2}
	\fmflabel{$q$}{od1}
	\fmflabel{$q$}{od2}
	\fmf{boson}{id1,vd1}
	\fmf{boson}{vd2,id2}
	\fmf{plain, label=$Q$, l.s=right}{vd2,vd1}
	\fmf{phantom}{od1,vd1}
	\fmf{phantom}{vd2,od2}
	\fmf{plain,tension=0}{od1,vd2}
	\fmf{plain,tension=0}{vd1,od2}
      \end{fmfgraph*}}
    \end{fmffile}
  \end{center}
  \caption{Vector dark matter scattering diagrams}
\end{figure}
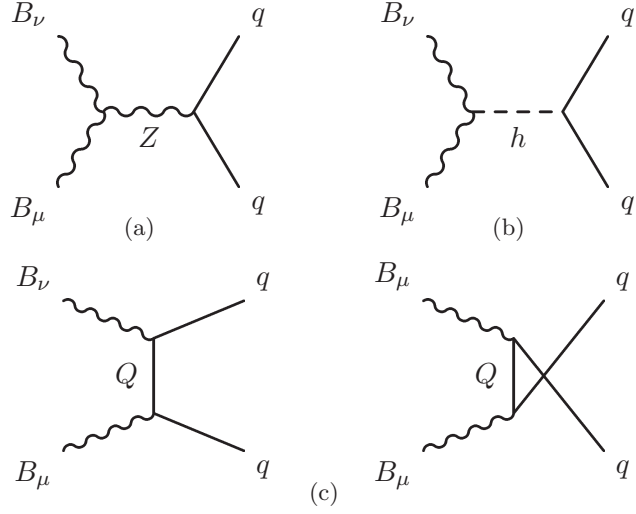

There are three classes of diagrams that can contribute to the scattering of 
vector dark matter with SM quarks at tree-level. Diagrams \ref{fig:v1b} and 
\ref{fig:v1c} arise in the context of universal extra 
dimensions~\cite{Cheng:2002ej,Servant:2002hb} and also in the context of the 
littlest Higgs with T-parity~\cite{Birkedal:2006fz}.

\subsubsection* {Diagram \ref{fig:v1a}: t-channel vector exchange}

An interaction of this form arises only if the vector field is complex. The 
reason is that the triple gauge boson coupling is proportional to the 
completely anti-symmetric structure constant $f^{abc}$, which requires the 
presence of at least three distinct real vector dark matter fields. For 
simplicity we assume that the interaction is mediated by a single real 
vector boson $Z$. Our results can easily be generalized to more complicated 
cases. We let the gauge index take values $1,2,3$, and denote the structure 
constant $f^{123}$ by $f$.  We define the dark matter field $B_\mu$ in 
terms of the $a=1,2$ components as
\begin{align}
	B_\mu
	&= \frac{A_\mu^1+iA_\mu^2}{\sqrt{2}}
	,
\end{align}
and associate the intermediate particle $Z_\mu$ with the gauge
index $a=3$.  We start from the following Lagrangian, dropping terms
which are irrelevant for this process.
\begin{align}
	\mathcal{L}
	&=
	-\frac14\mathcal{F}_{\mu\nu}^a\mathcal{F}^{\mu\nu a}
	+\frac12 m_Z^2 Z_\mu Z^\mu
	+g\bar{q}\gamma^\mu(\alpha-\beta\gamma^5) q Z_\mu
	\nonumber \\
	&\simeq
	-\frac14(\partial_\mu Z_\nu-\partial_\nu Z_\mu)^2 
	+\frac12 m_Z^2 Z_\mu Z^\mu 
	+g\bar{q}\gamma^\mu(\alpha-\beta\gamma^5) q Z_\mu
	\nonumber \\ 
	&\qquad+ \left[i g f
	\left(
	Z^\mu B^{\nu\dagger}
	(\partial_\mu B_\nu-\partial_\nu B_\mu)
	+B_\mu^\dagger B_\nu
	\partial^\mu Z^\nu
	\right)
	+ \text{h.c.}
	\right].
\end{align}
From the equation of motion for $Z_\mu$, keeping terms upto 
first order in $\partial /m_Z $, we find 
\begin{align}
  Z^\mu
  &\simeq
  -\frac{1}{m_Z^2}
  \left[
  g\bar{q}\gamma^\mu(\alpha-\beta\gamma^5) q 
  +igf(B_\nu^\dagger \partial^\mu B^\nu
  -B_\nu^\dagger \partial^\nu B^\mu
  -B_\nu \partial^\mu B^{\nu\dagger}
  +B_\nu \partial^\nu B^{\mu\dagger})
  \right.
  \nonumber \\
  &\left.\qquad\qquad
  -igf \partial^\nu
  (B_\nu^\dagger B^\mu-B^{\mu\dagger} B_\nu)
  \right]
  .
\end{align}
 Eliminating $Z_\mu$ using its equation of motion and keeping only the terms 
relevant for this process that survive in the non-relativistic limit, we 
obtain the effective Lagrangian
\begin{align}
  \mathcal{L}_{\rm{eff}}
  &\simeq
  \frac{i\alpha g^2 f}{m_Z^2}
  \left( B_\nu \partial_\mu B^{\nu\dagger}
  - \partial_\mu B_\nu B^{\nu\dagger} \right) 
  \;
  \bar{q}\gamma^\mu q
  .
\end{align}
This generates purely spin-independent interactions. 
\subsubsection* {Diagram \ref{fig:v1b}: t-channel scalar exchange}
We first start with the case of a real vector dark matter field. The relevant 
part of the 
Lagrangian has the general form
\begin{align}
  \mathcal{L}
  &= 
  \frac12 (\partial h)^2 - \frac12 m_h^2 h^2
  +a B_{\mu}B^{\mu} h
  -\bar{q}(\wt\alpha-\wt\beta\gamma^5)q h
  .
\end{align}
Integrating out $h$, and ignoring terms of order
$\partial^2/m_{h}^2$ and higher, we find
\begin{align}
  h &\simeq
  \frac{a B_{\mu}B^{\mu}
  -\bar{q}(\wt\alpha-\wt\beta\gamma^5)q
  }{m_h^2}
  .
\end{align}
The relevant terms in the effective Lagrangian are then
\begin{align}
  \mathcal{L}_{\rm{eff}}
  &\simeq
  -\frac{a\widetilde\alpha}
  {m_h^2}B_{\mu}B^{\mu}\bar{q}q
  .
\end{align}
This diagram generates spin-independent interactions that are suppressed in 
the chiral limit. It is straightforward to verify that in the case of 
complex vector dark matter, the conclusion is exactly the same. The 
Lagrangian we began from explicitly breaks chiral symmetry, which is why 
this result differs from our naive operator analysis.

\subsubsection* {Diagram \ref{fig:v1c}: s- and u-channel fermion exchange} 

We again start with the case that the dark matter field is real. The most 
general renormalizable Lagrangian for such a process takes the form
\begin{align}
  \mathcal{L}
  &=
  \bar{Q}(i\slashed{\partial}-m_Q)Q
  +\bar{q}\gamma^{\mu}(\alpha-\beta\gamma^5)Q\;B_{\mu}
  +\bar{Q}\gamma^{\mu}(\alpha^*-\beta^*\gamma^5)q\;B_{\mu}
  .
\end{align}
Integrating out $Q$ at tree-level we find
\begin{align}
  Q
  &=\frac{i\slashed{\partial}+m_Q}{\partial^2+m_Q^2}
  \left[
  \gamma^{\mu}(\alpha^*-\beta^*\gamma^5)q\;B_{\mu}
  \right]
  .
\end{align}
This leads to the effective Lagrangian
\begin{align}
  \mathcal{L}_{\rm eff}
  &=\bar{q}\gamma^{\nu}(\alpha-\beta\gamma^5)B_{\nu}
  \frac{i\slashed{\partial}+m_Q}{\partial^2+m_Q^2}
  \left[
  \gamma^{\mu}(\alpha^*-\beta^*\gamma^5)q\;B_{\mu}
  \right]
  .
\end{align}
Once again keeping only zeroth order and first order
terms in an expansion in powers of $\partial/m_Q$ we find
\begin{align}
  \mathcal{L}_{\rm eff}
  &=
  \frac{1}{m_Q^2}
  \left[
  m_Q \;
  \bar{q}\gamma^{\nu}(\alpha-\beta\gamma^5)
  \gamma^{\mu}(\alpha^*-\beta^*\gamma^5)q\;B_{\mu}
  B_{\nu}
  \right]
  \nonumber
  \\&\qquad
  +
  \frac{i}{m_Q^2}\left[
  \bar{q}\gamma^{\nu}(\alpha-\beta\gamma^5)
  \gamma^\alpha\gamma^{\mu}(\alpha^*-\beta^*\gamma^5)q\;
  B_{\nu}
  \partial_\alpha
  B_{\mu}
  \right]
  .
\end{align}
where we have neglected terms where the derivatives act on
the quark fields, and terms which vanish in the non-relativistic limit.
After integrating by parts and using the identity
$\gamma^\mu\gamma^\alpha\gamma^\nu-\gamma^\nu\gamma^\alpha\gamma^\mu
= 2i\epsilon^{\alpha\mu\nu\rho} \gamma_\rho\gamma^5$, this reduces to
\begin{align}
  \mathcal{L}_{\rm eff}
  &=
  \frac{1}{m_Q^2}
  \left[
  m_Q \;
  (|\alpha|^2-|\beta|^2)
  \bar{q} q\;B_{\mu}
  B^{\mu}
  \right]
  \nonumber
  \\&\qquad
  -\epsilon^{\mu\nu\rho \sigma}
  \frac{1}{m_Q^2}
  \left[
  (|\alpha|^2+|\beta|^2)
  \bar{q}
  \gamma_{\rho}
  \gamma^5
  q\;
  -(\alpha\beta^*+\beta\alpha^*)
  \bar{q}
  \gamma_{\rho}
  q\;
  \right]
  B_{\nu}
  \partial_\sigma
  B_{\mu}
  .
\end{align}
In the low energy limit, the derivative picks out the
$\sigma=0$ component in the second term. The $\epsilon$ tensor
then requires the other indices to be spatial. Therefore the term
proportional to the quark vector current is velocity suppressed.
The effective Lagrangian is then
\begin{align}
  \mathcal{L}_{\rm eff}
  &=
  \frac{1}{m_Q^2}
  \left[
  m_Q \;
  (|\alpha|^2-|\beta|^2)
  \bar{q} q\;B_{\mu}
  B^{\mu}
  -\epsilon^{\mu\nu\rho\sigma}
  (|\alpha|^2+|\beta|^2)
  \bar{q}
  \gamma_{\rho}
  \gamma^5
  q\;
  B_{\nu}
  \partial_\sigma
  B_{\mu}
  \right]
  .
\end{align}
The spin-independent contribution vanishes in the limit of exact chiral
symmetry, and can therefore naturally be small. Therefore, the
spin-dependent cross section naturally dominates in this case.

We now move on to the case where the vector dark matter field is complex. 
Then only the u-channel diagram contributes. The Lagrangian we start from
is modified to
\begin{align}
  \mathcal{L}
  &=
  \bar{Q}(i\slashed{\partial}-m_Q)Q
  +\bar{q}\gamma^{\mu}(\alpha-\beta\gamma^5)Q\;B_{\mu}
  +\bar{Q}\gamma^{\mu}(\alpha^*-\beta^*\gamma^5)q\;B_{\mu}^\dagger
  .
\end{align}
The equation of motion for $\bar{Q}$ leads to 
\begin{align}
  Q
  &=\frac{i\slashed{\partial}+m_Q}{\partial^2+m_Q^2}
  \left[
  \gamma^{\mu}(\alpha^*-\beta^*\gamma^5)q\;B_{\mu}^\dagger
  \right]
  .
\end{align}
The effective Lagrangian is then
\begin{align}
  \mathcal{L}_{\rm eff}
  &=\bar{q}\gamma^{\nu}(\alpha-\beta\gamma^5)B_{\nu}
  \frac{i\slashed{\partial}+m_Q}{\partial^2+m_Q^2}
  \left[
  \gamma^{\mu}(\alpha^*-\beta^*\gamma^5)q\;B_{\mu}^\dagger
  \right]
  .
\end{align}
Working to first order in $\partial/m_Q$, we are left with
\begin{align}
  \mathcal{L}_{\rm eff}
  &=\frac{B_\mu B_\nu^\dagger}{m_Q}
  \;\;\bar{q}\left[(|\alpha|^2-|\beta|^2)\gamma^\mu\gamma^\nu
  +(\alpha^*\beta-\alpha\beta^*)\gamma^\mu\gamma^\nu\gamma^5
  \right] q \nonumber \\
  &\qquad +\frac{i B_\mu \partial_\alpha B_\nu^\dagger}{m_Q^2}
  \;\;\bar{q}\left[(|\alpha|^2+|\beta|^2)\gamma^\mu\gamma^\alpha\gamma^\nu
  -(\alpha^*\beta+\alpha\beta^*)\gamma^\mu\gamma^\alpha\gamma^\nu\gamma^5
  \right] q
  .
\end{align}
Here we have neglected the action of derivatives on quark fields.  To 
distinguish spin-independent and spin-dependent contributions, we make use 
of the following identities.
\begin{align}
  \gamma^\mu\gamma^\nu
  &=
  g^{\mu\nu}-i\sigma^{\mu\nu} \nonumber \\
  \gamma^\mu\gamma^\nu\gamma^5
  &=
  g^{\mu\nu}\gamma^5
  +\frac12\epsilon^{\mu\nu\alpha\beta}\sigma_{\alpha\beta} 
  \nonumber \\
  \gamma^\mu\gamma^\alpha\gamma^\nu
  &=g^{\mu\alpha}\gamma^\nu
  -g^{\mu\nu}\gamma^\alpha
  +g^{\nu\alpha}\gamma^\mu
  +i\epsilon^{\mu\alpha\nu\rho}\gamma_\rho\gamma^5
  \nonumber \\
  \gamma^\mu\gamma^\alpha\gamma^\nu\gamma^5
  &=g^{\mu\alpha}\gamma^\nu\gamma^5
  -g^{\mu\nu}\gamma^\alpha\gamma^5
  +g^{\nu\alpha}\gamma^\mu\gamma^5
  +i\epsilon^{\mu\alpha\nu\rho}\gamma_\rho
  .
\end{align}
Dropping terms that do not survive in the non-relativistic limit, we find 
the form of the effective Lagrangian to be
\begin{align}
  \mathcal{L}_{\rm eff}
  &=
  \frac{|\alpha|^2-|\beta|^2}{m_Q}
  \left(
  B_\mu^\dagger B^\mu\;\bar{q}q
  -iB_\mu B_\nu^\dagger\;\bar{q}\sigma^{\mu\nu}q
  \right)\nonumber \\
  &\qquad+
  \frac{i(|\alpha|^2+|\beta|^2)}{m_Q^2}
  \left[
  -B^\mu(\partial_\alpha B_\mu^\dagger)
  \;\bar{q}\gamma^\alpha q
  +i\epsilon^{\mu\alpha\nu\rho}B_\mu(\partial_\alpha B_\nu^\dagger)
  \;\bar{q}\gamma_\rho\gamma^5q
  \right]
  .
\end{align}
We see that even in the limit of exact chiral symmetry, both 
spin-independent and spin-dependent interactions are present. This is 
exactly as expected, since there are no values of the parameters $\alpha$ 
and $\beta$ for which the theory admits a $B_\mu \leftrightarrow - 
{B_\mu}^{\dagger}$ interchange symmetry. It is not possible to incorporate 
such a symmetry in a manner consistent with the rephasing symmetry of 
$B_\mu$ unless additional fields are added to the theory.

In conclusion, we see that in the chiral limit, the results of scalar, real 
vector and Majorana fermion dark matter are in perfect agreement with the 
operator analysis we performed earlier.  Scalar dark matter always leads to 
spin-independent interactions, while in the limit of exact chiral symmetry 
Majorana fermions and real vector bosons give rise to spin-dependent 
interactions. There is always a sizeable spin-independent component to the 
WIMP-nucleon cross-section in the cases of Dirac fermion and complex vector boson 
dark matter, except for very specific choices of quantum numbers. From this it 
follows that primarily spin-dependent WIMP-nucleon cross sections are closely 
associated with theories where the dark matter particle is its own anti-particle.

\section{Models with spin-dependent couplings}

We now study in more detail the cases which naturally lead to purely
spin-dependent interactions in the chiral limit. We continue to work in
the non-relativistic regime, and under the assumption that the momentum
transfer in the scattering process is much less than both the dark matter
mass and the characteristic nuclear scales. We however no longer assume
that the dark matter mass is much less than the mass of the particle
mediating the interaction.

\subsection{Dirac fermion}
\subsubsection* {t-channel vector exchange}
\begin{figure}[h]
  \begin{center}
    \begin{fmffile}{d}
      \label{fig:d}
      \begin{fmfgraph*}(30,20)
        \fmfleftn{ia}{2} \fmfrightn{oa}{2}
        \fmflabel{$\chi,\chi^c$}{ia1}
        \fmflabel{$\chi,\chi^c$}{ia2}
        \fmflabel{$q$}{oa1}
        \fmflabel{$q$}{oa2}
        \fmf{plain}{ia1,va1}
        \fmf{plain}{va1,ia2}
        \fmf{boson, label=$Z$}{va1,va2}
        \fmf{plain}{oa1,va2}
        \fmf{plain}{va2,oa2}
      \end{fmfgraph*}
    \end{fmffile}
  \end{center}
  \caption{Dirac dark matter scattering through t-channel vector exchange}
\end{figure}

We return to the Lagrangian
\begin{align}
  \mathcal{L}&=
  -\frac{1}{4}\mathcal{F}^{\mu\nu}\mathcal{F}_{\mu\nu}
  +\frac{1}{2}m_Z^2\;Z^{\mu}Z_{\mu}
  +\bar{\chi}\gamma^{\mu}
  (\alpha-\beta\gamma^5)\chi Z_{\mu}
  +\bar{q}\gamma^{\mu}(\widetilde\alpha-\widetilde\beta\gamma^5)q Z_{\mu}
  .
\end{align}
Since the momentum in the propagator in the case of t-channel exchange
is negligible compared to the dark matter mass, we can continue to use
the effective Lagrangian defined in equation (\ref{eq:leffdfa}),
\begin{align}
  \mathcal{L}_{\rm eff}
  &\simeq
  -\frac{1}{m_Z^2}
  \left[
  \alpha\widetilde\alpha\;
  \bar{\chi}\gamma^{\mu}\chi \;
  \bar{q}\gamma_{\mu}q
  +\beta\widetilde\beta \;
  \bar{\chi}\gamma^{\mu}\gamma^5\chi \;
  \bar{q}\gamma_{\mu}\gamma^5q
  \right]
  .
\end{align}
This leads to purely spin-dependent scattering only in the cases when
either $\alpha$ or $\widetilde{\alpha}$ is zero, while both $\beta$ and
$\widetilde{\beta}$ are non-zero.  

Consider first the possibility that $\widetilde{\alpha} = 0$. Clearly the 
charges of the quarks under the SM $Z$ are such that this criterion is not 
satisfied. The crucial question is then whether a $Z'$ can exist for which 
such a charge assignment is phenomenologically viable. The requirement 
that $\widetilde{\beta} \neq 0$ implies that the left- and right-handed 
quarks have different charges under the $Z'$. This in turn means that 
either the SM Higgs is charged under the $Z'$, or alternatively that the 
quark masses arise from non-renormalizable interactions. We consider each 
of these possibilities in turn.

If the SM Higgs is charged under the $Z'$, when electroweak symmetry is 
broken the SM $Z$ will mix with the $Z'$. The mixing angle is of order 
$(m_Z/m_Z')^2$. After diagonalization the dark matter particle will be 
acquire a charge under the $Z$ of order $(m_Z/m_Z')^2$. This will generate 
a sizeable spin-independent contribution to the WIMP-nucleon cross 
section. We conclude that this approach is not viable.

If the SM Higgs is not charged under the $Z'$, then the quark masses must 
arise from non-renormalizable interactions involving both the SM Higgs and 
the Higgs field $H'$ that breaks the gauge symmetry corresponding to the 
$Z'$.  While this is perhaps adequate for the light quarks, it is somewhat 
unsatisfactory for the top quark, whose mass is large. This problem can be 
avoided by assigning the three generations different charges under the 
$Z'$, but at the expense of a new source of flavor violation close to the 
weak scale. In addition there is a chirality suppressed spin-independent 
contribution to the WIMP-nucleon cross section from $H'$ exchange, which 
may be sizeable. We conclude that this scenario is disfavored, although 
perhaps not excluded.

We now consider the alternate scenario where $\alpha =0$. If the Dirac fermion 
$\chi$ has no interactions beyond those in the Lagrangian above, a simple change 
of variables allows the theory with $\alpha = 0$ to be rewritten as a theory of 
two degenerate Majorana fermions with identical charges under $Z$. This theory is 
therefore not distinct from the case of Majorana fermion dark matter to be 
considered in the next section. If $\chi$ does have additional interactions, 
however, this theory of Dirac fermion dark matter is in general not equivalent to 
a theory of Majorana fermion dark matter. We conclude that this class of theories 
of Dirac fermion dark matter can indeed give rise to primarily spin-dependent 
WIMP-nucleon scattering at tree-level, but only if the WIMP carries very specific 
charges under the gauge boson. We will not consider this possibility further.

\subsection{Majorana fermion}
\subsubsection* {t-channel vector exchange}
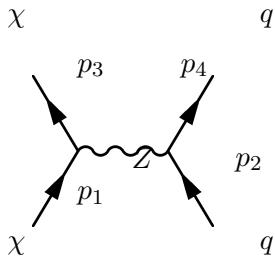
\begin{figure}[h]
  \begin{center}
    \begin{fmffile}{ma}
      \begin{fmfgraph*}(30,20)
	\fmfleftn{ia}{2} \fmfrightn{oa}{2}
	\fmflabel{$\chi$}{ia1}
	\fmflabel{$\chi$}{ia2}
	\fmflabel{$q$}{oa1}
	\fmflabel{$q$}{oa2}
	\fmf{fermion, label=$p_1$}{ia1,va1}
	\fmf{fermion, label=$p_3$}{va1,ia2}
	\fmf{boson, label=$Z$}{va1,va2}
	\fmf{fermion,label=$p_2$}{oa1,va2}
	\fmf{fermion,label=$p_4$}{va2,oa2}
      \end{fmfgraph*}
    \end{fmffile}
  \end{center}
  \caption{Majorana dark matter scattering through t-channel vector 
exchange}
  \label{fig:ma}
\end{figure}
We start from the Lagrangian
\begin{align}
  \mathcal{L}&=
  -\frac{1}{4}\mathcal{F}^{\mu\nu}\mathcal{F}_{\mu\nu}
  +\frac{1}{2}m_Z^2\;Z^{\mu}Z_{\mu}
  +\bar{\chi}\gamma^{\mu}(-\beta\gamma^5)\chi Z_{\mu}
  +\bar{q}\gamma^{\mu}(\widetilde{\alpha}-\widetilde\beta\gamma^5)qZ_{\mu}
  .
\end{align}
We can again neglect momentum-dependent terms in the propagator, leading
to the matrix element
\begin{align}
  \mathcal{M}
  &=
  -\frac{2\beta\widetilde\beta}{m_Z^2}
  \bar{u}_\chi \gamma^\mu\gamma^5u_\chi
  \langle \bar q\gamma_\mu\gamma^5 q\rangle
  .
\end{align}

The matrix elements $\langle \bar{q} \gamma_{\mu} \gamma^5 q \rangle$ etc. 
are defined in Appendix \ref{sec:low-energy}, while $u_\chi,\bar{u}_\chi$ 
are the familiar spinors corresponding to the dark matter fermion. We see 
that this diagram does indeed lead to purely spin-dependent cross 
sections. The gauge boson exchanged in this process may be either 
the SM $Z$, 
or a new $Z'$. If it is the SM $Z$, then the dark matter particle must 
either constitute the neutral component of a single representation of SM 
SU(2$)_{\rm L}$ or arise as a linear combination of the neutral components 
of different SU(2$)_{\rm L}$ representations. There is no such constraint 
if the gauge boson is a $Z'$. The physics of these two cases is therefore 
very different, and so we consider them separately.

\medskip
\noindent
\underline{Exchange of the Standard Model Z}

Consider first the case of the SM $Z$. Since the $Z$ is light and all its 
couplings are fixed, this scenario is already somewhat constrained by 
experiments. Let us understand the nature of these bounds. As explained 
above, the dark matter must either be the neutral component of an 
SU(2$)_{\rm L}$ representation, or a linear combination of the neutral 
components of different SU(2$)_{\rm L}$ representations and SM singlets. 
If the dark matter is not part of a linear combination, the direct 
detection constraints arising from spin-dependent WIMP-nucleon 
interactions are very strong. For example, a Majorana neutrino in a pure 
SU(2$)_{\rm L}$ doublet representation has been excluded by XENON as a 
dark matter candidate \cite{Angle:2008we}. For this reason we expect that 
in this class of theories the WIMP will be a linear combination of the 
neutral components of different SU(2$)_{\rm L}$ representations, which 
allows the possibility of weaker couplings to the $Z$, thereby loosening
the direct detection constraints.

In addition to this spin-dependent contribution to the WIMP-nucleon cross 
section from $Z$ exchange, there is also necessarily a spin-independent 
contribution from SM Higgs exchange. We now explain the origin of this 
effect. The coupling of any chiral field in an SU(2$)_{\rm L}$ 
representation to the $Z$ is proportional to $I_3 + Q\sin^2 \theta_W $, 
where $I_3$ is the weak-isospin, $Q$ is the electric charge and $\theta_W$ 
is the weak mixing angle. It follows that for the neutral component to 
have non-zero charge under the $Z$, the representation must carry 
hypercharge. Therefore, until electroweak symmetry is broken, it cannot 
acquire a Majorana mass. This implies that the Majorana mass of any dark 
matter particle charged under the $Z$ can arise only as an electroweak 
symmetry breaking effect, from couplings involving the Higgs. The 
conclusion is that in this class of theories there is a spin-independent 
contribution to the WIMP-nucleon cross section mediated by the Higgs, 
which is in general correlated with the spin-dependent cross section 
mediated by the $Z$~\cite{Cohen:2010gj}.

In order to obtain a quantitative understanding of the constraints on this 
scenario, we choose a benchmark model. We consider a theory where the dark 
matter particle arises as a linear combination of the neutral components 
of two SM SU(2$)_{\rm L}$ doublets which have hypercharges 
${\rm Y} = \pm \frac12$.
\begin{align}
  \xi=
  \left(
  \begin{array}{c}
    \xi^+\\
    \xi_0
  \end{array}
  \right)
  \qquad
  \xi^c=
  \left(
  \begin{array}{c}
    \xi^c_0\\
    \xi^{c-}
  \end{array}
  \right)
  .
\end{align}
In addition to a Dirac mass for $\xi$ and $\xi^c$, we include a 
non-renormalizable operator that gives rise to a Majorana 
mass term for the neutral component of $\xi$. 
\begin{align}
  \mathcal{L} \supset -\frac{(H^\dagger \xi)^2}{\Lambda}
  +{\rm h.c.}
\end{align}
Here $H$ is the SM Higgs doublet.
This non-renormalizable operator can be generated by integrating out a
SM singlet.
The Higgs acquires a vacuum expectation value, 
\begin{align}
 \langle H \rangle &=
 \frac{1}{\sqrt{2}}
  \left(
  \begin{array}{c}
    0\\
    v
  \end{array}
  \right)
  ,
\end{align}
where $v =$ 246 GeV.
We generate a Majorana mass term and a Yukawa
coupling for the neutral component of the field $\xi_0$.
We denote the physical Higgs field of the SM that emerges after 
electroweak symmetry breaking
by $h$. 
The Lagrangian now contains,
\begin{align}
  \mathcal{L} &\supset 
  -
  \frac{g}{\cos\theta_W}
  \left(\bar{\xi}_0 \frac12 \bar{\sigma}^\mu \xi_0
  -\bar{\xi}_0^c \frac12 \bar{\sigma}^\mu \xi_0^c\right)
  Z_\mu
  \nonumber
  \\&\qquad\qquad
  -
  \left[
  \frac12
  \left(
  \begin{array}{cc}
    \xi^c_0 & \xi_{0}
  \end{array}
  \right)
  \left(
  \begin{array}{cc}
    0 & M\\
    M & m
  \end{array}
  \right)
  \left(
  \begin{array}{c}
    \xi^c_0\\
    \xi_{0}
  \end{array}
  \right)
  +
  y_\xi \xi_0 \xi_0 h
  +{\rm h.c.}
  \right]
  ,
\end{align}
where $m = {v^2}/{\Lambda}$ and $y_\xi =  m/v$.
The lighter eigenstate of the two mass eigenstates $\xi_D$ is the dark 
matter field,
\begin{align}
  \xi_D = 
  \cos \phi\; \xi_0 
  +\sin\phi\; \xi_0^c 
  .
\end{align}
where $\phi$ is the mixing angle. 
The couplings of $\xi_D$ in the mass basis are given by
\begin{align}
  \mathcal{L} \supset 
  -\frac{g\cos2\phi}{2\cos\theta_W}
  \bar{\xi}_D\bar{\sigma}^\mu \xi_D
  Z_\mu
  -
  \left[
  y_\xi\cos^2 \phi\; \xi_D \xi_D\,h
  +{\rm h.c.}
  \right]
  .
\end{align}
Translating this to couplings with a four-component Majorana
fermion, $\chi$, to be consistent with the rest of our analysis, we
find,
\begin{align}
  \mathcal{L} \supset 
  \frac{g\cos2\phi}{4\cos\theta_W}
  \bar{\chi}\gamma^\mu \gamma^5 \chi
  Z_\mu
  -
  y_\xi \cos^2\phi \;\bar{\chi} \chi h
  .
\end{align}

The coupling to the $Z$ is suppressed by $\cos 2\phi$. In the limit $m\ll 
M$, this is simply $-m/2M$. The dark matter mass in this limit is 
approximately equal to $M$. Therefore, we see that at higher dark matter 
masses the XENON spin-dependent bounds can be avoided. However, since the 
coupling of the WIMP to the Higgs $y_{\xi}$ is proportional to the 
Majorana mass term $m$ and independent of $M$, the spin-independent cross 
section is insensitive to the dark matter mass in this limit. Therefore 
the spin-dependent and spin-independent bounds are somewhat complementary, 
with the latter more effective for larger values of the dark matter mass.

\begin{figure}[htp]
  \begin{center}
    \includegraphics{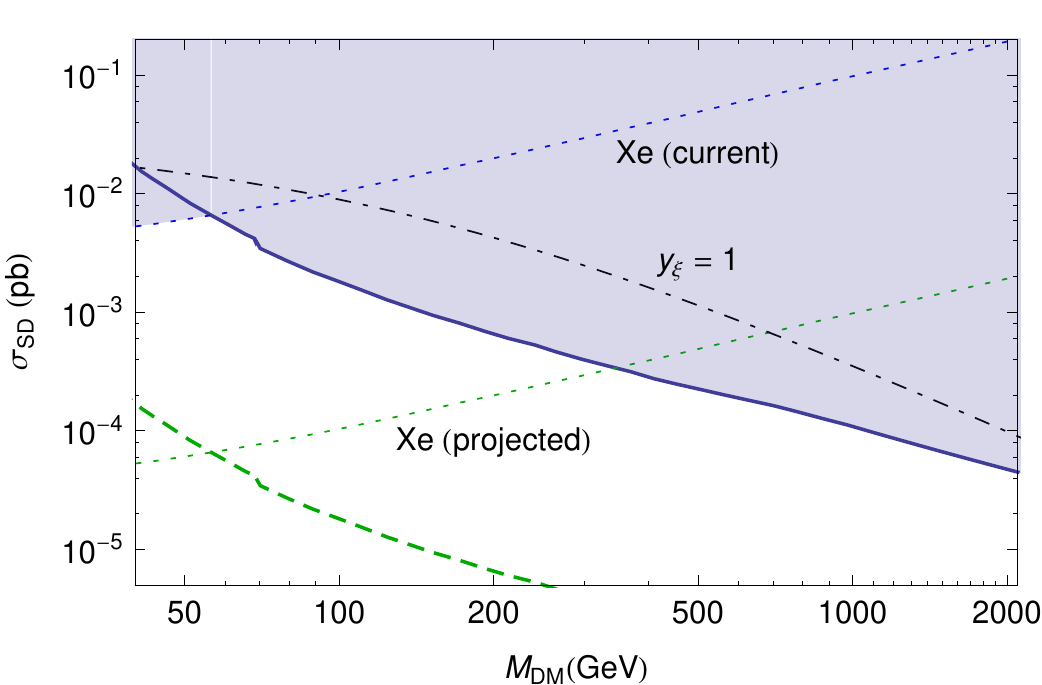}
  \end{center}
  \caption{The maximum spin-dependent WIMP-nucleon cross section in the
  benchmark model consistent with the current (blue solid) and projected 
  (green dashed) bounds on the spin-independent cross section from direct
  detection experiments. Also shown are
  current (blue dotted) and projected (green dotted) bounds on
  spin-dependent scattering from
  the XENON experiment. The black dot-dashed line shows
  where the WIMP-Higgs Yukawa coupling $y_\xi = 1$.}
  \label{fig:zsd}
\end{figure}
In figure \ref{fig:zsd} we display the interplay between spin-dependent 
and spin-independent direct detection bounds in the benchmark model. The 
upper (lower) curve with positive slope shows the current (future) bounds 
on spin-dependent scattering from the XENON experiment as a function of 
the dark matter mass. The upper (lower) curve with negative slope shows 
the maximum value of the spin-dependent scattering cross section that is 
consistent with the current (future) spin-independent direct detection 
constraints, arising from the Higgs exchange contribution as explained 
above. In obtaining this bound we have assumed a Higgs mass of 120 GeV.
Note that the shaded blue region has already been excluded, either 
by the limits on the spin-dependent WIMP-nucleon cross section from XENON, 
or from spin-independent direct detection constraints. Therefore, in the 
near future, spin-dependent searches can only expect to find a signal in 
the triangular region on the left side of the plot, at dark matter masses 
below about 400 GeV. In most of this region spin-independent searches also 
have discovery potential. 
We have also included in this figure a curve that indicates the value of 
the spin-dependent cross section corresponding to a Yukawa coupling of 1 
as a function of the dark matter mass, to establish that the entire region 
of interest for upcoming experiments can be studied reliably using 
perturbation theory. Although these conclusions apply strictly only to 
this benchmark model, we expect that similar results will hold in the 
more general case of Majorana fermion dark matter scattering mediated by 
the SM $Z$.

\medskip
\noindent
\underline{Exchange of a New $Z'$}

If the gauge boson being exchanged is not the SM $Z$ but a new $Z'$,
there is considerably more flexibility with regard to charge
assignments.  However, the severe constraints from precision
electroweak measurements, direct
production and four-fermion point interactions~\cite{Appelquist:2002mw,Amsler:2008zzb} 
on the mass and
couplings of any new gauge boson limits the possible signal in direct
detection experiments. In addition to the spin-dependent contribution
to the WIMP-nucleon cross section, there may also be a
spin-independent contribution if the SM Higgs mixes with the Higgs
that gives mass to the $Z'$. However, whether this effect arises or
not is dependent on the $Z'$ charge assignments, and therefore does
not give rise to a robust bound.  This is another difference from the
case of the SM $Z$.

\subsubsection* {s- and u-channel vector exchange}
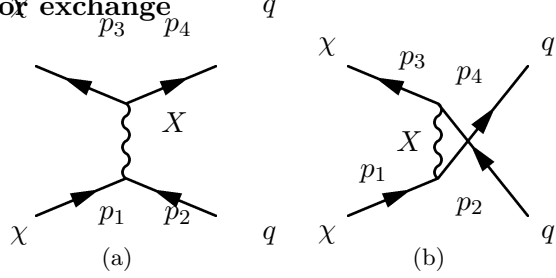
\begin{figure}[h]
  \begin{center}
    \begin{fmffile}{mb}
      \subfloat[]{
      \label{fig:mba}
      \begin{fmfgraph*}(30,20)
	\fmfleftn{ix}{2} \fmfrightn{ox}{2}
	\fmflabel{$\chi$}{ix1}
	\fmflabel{$\chi$}{ix2}
	\fmflabel{$q$}{ox1}
	\fmflabel{$q$}{ox2}
	\fmf{fermion, label=$p_1$}{ix1,vx1}
	\fmf{fermion, label=$p_3$}{vx2,ix2}
	\fmf{boson, label=$X$}{vx1,vx2}
	\fmf{fermion, label=$p_2$}{ox1,vx1}
	\fmf{fermion, label=$p_4$}{vx2,ox2}
      \end{fmfgraph*}}
      \hspace{0.3in}
      \subfloat[]{
      \label{fig:mbb}
      \begin{fmfgraph*}(30,20)
	\fmfleftn{iy}{2} \fmfrightn{oy}{2}
	\fmflabel{$\chi$}{iy1}
	\fmflabel{$\chi$}{iy2}
	\fmflabel{$q$}{oy1}
	\fmflabel{$q$}{oy2}
	\fmf{fermion, label=$p_1$}{iy1,vy1}
	\fmf{fermion, label=$p_3$}{vy2,iy2}
	\fmf{boson, label=$X$}{vy1,vy2}
	\fmf{phantom,label=$p_2$,l.d=0.01mm,l.s=left}{oy1,vy1}
	\fmf{phantom,label=$p_4$,l.d=0.01mm,l.s=left}{vy2,oy2}
	\fmf{fermion,tension=0}{oy1,vy2}
	\fmf{fermion,tension=0}{vy1,oy2}
      \end{fmfgraph*}}
    \end{fmffile}
  \end{center}
  \caption{Majorana dark matter scattering through s- and u-channel vector 
           exchange}
  \label{fig:mb}
\end{figure}
The Lagrangian corresponding to this process is
\begin{align}
  \mathcal{L}
  &=
  -\frac{1}{2}\left|\partial_\mu X_\nu - \partial_\nu X_\mu\right|^2
  +m_X^2\;X^\dagger_\mu X^\mu
  +\bar{\chi}\gamma^{\mu}(\alpha-\beta\gamma^5)q
  X_{\mu}
  +\bar{q}\gamma^{\mu}(\alpha^*-\beta^*\gamma^5)\chi
  X_{\mu}^\dagger
  .
\end{align}
The matrix element for this process may be obtained from the Feynman
diagrams. After a Fierz rearrangement, it reduces to
\begin{align}
  \mathcal{M}
  &\simeq
  \frac{2}{(m_X^2-m_\chi^2)}
  \left[
  (|\alpha|^2-|\beta|^2)
  \langle \bar{q}q \rangle
  \bar{u}_\chi u_\chi
  -\frac{1}{2}
  (|\alpha|^2+|\beta|^2)
  \langle\bar{q}\gamma^{\mu}\gamma^{5}q\rangle
  \bar{u}_\chi \gamma_{\mu}\gamma^{5}u_\chi
  \right]
  .
\end{align}
We see that in the chiral limit, scattering is purely spin-dependent,
as expected. Note that the vector particle being exchanged is a
SM color triplet.

\subsubsection* {s- and u-channel scalar exchange}
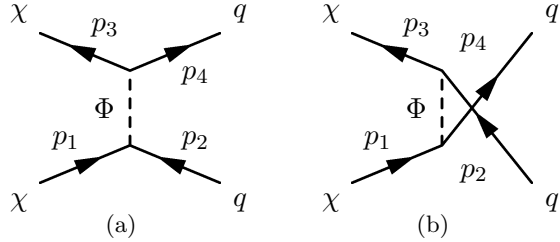
\begin{figure}[h]
  \begin{center}
    \begin{fmffile}{mc}
      \subfloat[]{
      \label{fig:mca}
      \begin{fmfgraph*}(30,20)
	\fmfleftn{ic}{2} \fmfrightn{oc}{2}
	\fmflabel{$\chi$}{ic1}
	\fmflabel{$\chi$}{ic2}
	\fmflabel{$q$}{oc1}
	\fmflabel{$q$}{oc2}
	\fmf{fermion, label=$p_1$}{ic1,vc1}
	\fmf{fermion, label=$p_3$}{vc2,ic2}
	\fmf{dashes, label=$\Phi$}{vc1,vc2}
	\fmf{fermion, label=$p_2$}{oc1,vc1}
	\fmf{fermion, label=$p_4$}{vc2,oc2}
      \end{fmfgraph*}}
      \hspace{0.3in}
      \subfloat[]{
      \label{fig:mcb}
      \begin{fmfgraph*}(30,20)
	\fmfleftn{id}{2} \fmfrightn{od}{2}
	\fmflabel{$\chi$}{id1}
	\fmflabel{$\chi$}{id2}
	\fmflabel{$q$}{od1}
	\fmflabel{$q$}{od2}
	\fmf{fermion, label=$p_1$}{id1,vd1}
	\fmf{fermion, label=$p_3$}{vd2,id2}
	\fmf{dashes, label=$\Phi$}{vd1,vd2}
	\fmf{phantom, label=$p_2$,l.s=left,l.d=0.01mm}{od1,vd1}
	\fmf{phantom, label=$p_4$,l.s=left,l.d=0.01mm}{vd2,od2}
	\fmf{fermion,tension=0}{od1,vd2}
	\fmf{fermion,tension=0}{vd1,od2}
      \end{fmfgraph*}}
    \end{fmffile}
  \end{center}
  \caption{Majorana dark matter scattering through s- and u-channel scalar exchange}
  \label{fig:mc}
\end{figure}
The Lagrangian corresponding to this process is
\begin{align}
  \mathcal{L}
  &=
  |\partial \Phi|^2-m_\Phi^2|\Phi|^2
  -\bar{\chi}(\alpha-\beta\gamma^5)q \Phi
  -\bar{q}(\alpha^*+\beta^*\gamma^5)\chi \Phi^{\dagger}
  .
\end{align}
After using Fierz identities and ignoring terms which are velocity 
suppressed, we can write the amplitude in terms of the expectation values 
of quark currents in the nucleus.
\begin{align}
  \mathcal{M}
  &=
  -\frac{1}{2(m_\Phi^2-m_\chi^2)}
  \left[
  \left(|\alpha|^2-|\beta|^2\right)
  \bar{u}_\chi u_\chi
  \langle \bar{q}q\rangle
  -
  \left(|\alpha|^2+|\beta|^2\right)
  \bar{u}_\chi\gamma^\mu\gamma^5 u_\chi
  \langle \bar{q}\gamma_\mu \gamma^5 q\rangle
  \right]
  .
\end{align}
As expected the cross section is purely spin-dependent in the chiral limit.
Again, note that the scalar being exchanged in this process transforms
as a fundamental under SM color.

\subsection{Real vector boson}
\subsubsection*{s- and u-channel fermion exchange}
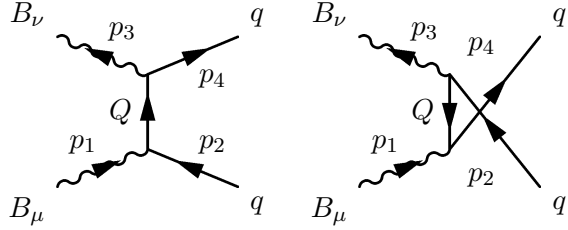
\begin{figure}[htp]
  \begin{center}
    \begin{fmffile}{v}
      \label{fig:v}
      \begin{fmfgraph*}(30,20)
	\fmfleftn{ic}{2} \fmfrightn{oc}{2}
	\fmflabel{$B_\mu$}{ic1}
	\fmflabel{$B_\nu$}{ic2}
	\fmflabel{$q$}{oc1}
	\fmflabel{$q$}{oc2}
	\fmf{boson}{ic1,vc1}
	\fmf{phantom_arrow, label=$p_1$, tension=0}{ic1,vc1}
	\fmf{boson}{ic2,vc2}
	\fmf{phantom_arrow, label=$p_3$, tension=0}{vc2,ic2}
	\fmf{fermion,label=$Q$}{vc1,vc2}
	\fmf{fermion, label=$p_2$}{oc1,vc1}
	\fmf{fermion, label=$p_4$}{vc2,oc2}
      \end{fmfgraph*}
      \hspace{0.3in}
      \begin{fmfgraph*}(30,20)
	\fmfleftn{id}{2} \fmfrightn{od}{2}
	\fmflabel{$B_\mu$}{id1}
	\fmflabel{$B_\nu$}{id2}
	\fmflabel{$q$}{od1}
	\fmflabel{$q$}{od2}
	\fmf{boson}{id1,vd1}
	\fmf{boson}{vd2,id2}
	\fmf{phantom_arrow, label=$p_1$, tension=0}{id1,vd1}
	\fmf{phantom_arrow, label=$p_3$, tension=0}{vd2,id2}
	\fmf{fermion, label=$Q$,l.s=right}{vd2,vd1}
	\fmf{phantom, label=$p_2$,l.s=left,l.d=0.01mm}{od1,vd1}
	\fmf{phantom, label=$p_4$,l.s=left,l.d=0.01mm}{vd2,od2}
	\fmf{fermion,tension=0}{od1,vd2}
	\fmf{fermion,tension=0}{vd1,od2}
      \end{fmfgraph*}
    \end{fmffile}
  \end{center}
  \caption{Vector dark matter scattering through s- and u-channel fermion exchange}
\end{figure}
The only diagram seen to give spin-dependent scattering was diagram
\ref{fig:v1c}. The relevant Lagrangian takes the form
\begin{align}
  \mathcal{L}
  &=
  \bar{Q}(i\slashed{\partial}-m_Q)Q
  +\bar{q}\gamma^{\mu}(\alpha-\beta\gamma^5)Q\;B_{\mu}
  +\bar{Q}\gamma^{\mu}(\alpha^*-\beta^*\gamma^5)q\;B_{\mu}
  .
\end{align}
The corresponding matrix element is
\begin{align}
  \mathcal{M}
  &=
  -\frac{2\epsilon_\mu(p_1)\epsilon^*_\nu(p_3)}
  {m_B^2-m_Q^2}
  \left[
  m_Q (|\alpha|^2-|\beta|^2) g^{\mu\nu}
  \langle\bar{q} q\rangle
  -
  i(|\alpha|^2+ |\beta|^2)
  m_B\epsilon^{0\mu\nu\rho}
  \langle\bar{q}\gamma_\rho\gamma^5 q\rangle
  \right]
  .
\end{align}
This is purely spin-dependent in the chiral limit. The fermion being
exchanged in this process is again a SM color triplet.

\begin{table}
  \begin{minipage}[htp]{\textwidth}
    \renewcommand{\thempfootnote}{\fnsymbol{mpfootnote}}
    \begin{center}
      \begin{tabular}{ |>{\centering}m{1.4in}| >{\centering}m{0.6in}|
	>{\centering}m{1.5in}|c|}
	\hline
	Dark Matter&Mediator & Process & Scattering\\
	\hline
	\multirow{3}{*}{Scalar}
	&$Z,Z'$& 
	\rule{0pt}{25pt}
	\begin{fmffile}{tab-s1}
	  \begin{fmfgraph*}(12,8)
	    \fmfleftn{ia}{2} \fmfrightn{oa}{2}
	    \fmf{dashes}{ia1,va1,ia2}
	    \fmf{boson}{va1,va2}
	    \fmf{plain}{oa1,va2,oa2}
	  \end{fmfgraph*}
	\end{fmffile}
	&
	SI\\
	\cline{2-4}
	&$h$
	&
	\rule{0pt}{25pt}
	\begin{fmffile}{tab-s2}
	  \begin{fmfgraph*}(12,8)
	    \fmfleftn{ib}{2} \fmfrightn{ob}{2}
	    \fmf{dashes}{ib1,vb1,ib2}
	    \fmf{dashes}{vb1,vb2}
	    \fmf{plain}{ob1,vb2,ob2}
	  \end{fmfgraph*}
	\end{fmffile}
	&  SI\\
	\cline{2-4}
	&$Q$&
	\rule{0pt}{25pt}
	\begin{fmffile}{tab-s3}
	  \begin{fmfgraph*}(12,8)
	    \fmfleftn{id}{2} \fmfrightn{od}{2}
	    \fmf{phantom}{id1,vd1}
	    \fmf{phantom}{id2,vd2}
	    \fmf{dashes,tension=0}{id2,vd1}
	    \fmf{dashes,tension=0}{id1,vd2}
	    \fmf{plain}{vd1,vd2}
	    \fmf{plain}{od1,vd1}
	    \fmf{plain}{vd2,od2}
	  \end{fmfgraph*}
	  ,
	  \begin{fmfgraph*}(12,8)
	    \fmfleftn{id}{2} \fmfrightn{od}{2}
	    \fmf{dashes}{id1,vd1}
	    \fmf{dashes}{id2,vd2}
	    \fmf{plain}{vd1,vd2}
	    \fmf{plain}{od1,vd1}
	    \fmf{plain}{vd2,od2}
	  \end{fmfgraph*}
	\end{fmffile}
	&  SI\\
	\hline
	\multirow{4}{*}{Dirac Fermion}
	&$Z$,$Z'$& 
	\rule{0pt}{25pt}
	\begin{fmffile}{tab-f1}
	  \begin{fmfgraph*}(12,8)
	    \fmfleftn{ia}{2} \fmfrightn{oa}{2}
	    \fmf{plain}{ia1,va1,ia2}
	    \fmf{boson}{va1,va2}
	    \fmf{plain}{oa1,va2,oa2}
	  \end{fmfgraph*}
	\end{fmffile}
	&  SI, SD\footnote[2]
	{Can be primarily SD for specific choices of $Z'$
	charges}
	\\
	\cline{2-4}
	&$h$
	&
	\rule{0pt}{25pt}
	\begin{fmffile}{tab-f2}
	  \begin{fmfgraph*}(12,8)
	    \fmfleftn{ib}{2} \fmfrightn{ob}{2}
	    \fmf{plain}{ib1,vb1,ib2}
	    \fmf{dashes}{vb1,vb2}
	    \fmf{plain}{ob1,vb2,ob2}
	  \end{fmfgraph*}
	\end{fmffile}
	&  SI\\
	\cline{2-4}
	&$X$&
	\rule{0pt}{25pt}
	\begin{fmffile}{tab-f3}
	  \begin{fmfgraph*}(12,8)
	    \fmfleftn{ic}{2} \fmfrightn{oc}{2}
	    \fmf{plain}{ic1,vc1}
	    \fmf{plain}{vc2,ic2}
	    \fmf{boson,tension=0.5}{vc1,vc2}
	    \fmf{phantom}{oc1,vc1}
	    \fmf{phantom}{vc2,oc2}
	    \fmf{plain,tension=0}{oc1,vc2}
	    \fmf{plain,tension=0}{vc1,oc2}
	  \end{fmfgraph*}
	  ,
	  \begin{fmfgraph*}(12,8)
	    \fmfleftn{id}{2} \fmfrightn{od}{2}
	    \fmf{plain}{id1,vd1}
	    \fmf{plain}{id2,vd2}
	    \fmf{boson,tension=0.5}{vd1,vd2}
	    \fmf{plain}{od1,vd1}
	    \fmf{plain}{vd2,od2}
	  \end{fmfgraph*}
	\end{fmffile}
	&  SI, SD\\
	\cline{2-4}
	&$\Phi$&
	\rule{0pt}{25pt}
	\begin{fmffile}{tab-f4}
	  \begin{fmfgraph*}(12,8)
	    \fmfleftn{id}{2} \fmfrightn{od}{2}
	    \fmf{plain}{id1,vd1}
	    \fmf{plain}{vd2,id2}
	    \fmf{dashes}{vd1,vd2}
	    \fmf{phantom}{od1,vd1}
	    \fmf{phantom}{vd2,od2}
	    \fmf{plain,tension=0}{od1,vd2}
	    \fmf{plain,tension=0}{vd1,od2}
	  \end{fmfgraph*}
	  ,
	  \begin{fmfgraph*}(12,8)
	    \fmfleftn{id}{2} \fmfrightn{od}{2}
	    \fmf{plain}{id1,vd1}
	    \fmf{plain}{id2,vd2}
	    \fmf{dashes}{vd1,vd2}
	    \fmf{plain}{od1,vd1}
	    \fmf{plain}{vd2,od2}
	  \end{fmfgraph*}
	\end{fmffile}
	&  SI, SD\\
	\hline
	\multirow{4}{*}{Majorana Fermion}
	&$Z$,$Z'$& 
	\rule{0pt}{25pt}
	\begin{fmffile}{tab-fm1}
	  \begin{fmfgraph*}(12,8)
	    \fmfleftn{ia}{2} \fmfrightn{oa}{2}
	    \fmf{plain}{ia1,va1,ia2}
	    \fmf{boson}{va1,va2}
	    \fmf{plain}{oa1,va2,oa2}
	  \end{fmfgraph*}
	\end{fmffile}
	&  SD\\
	\cline{2-4}
	&$h$
	&
	\rule{0pt}{25pt}
	\begin{fmffile}{tab-fm2}
	  \begin{fmfgraph*}(12,8)
	    \fmfleftn{ib}{2} \fmfrightn{ob}{2}
	    \fmf{plain}{ib1,vb1,ib2}
	    \fmf{dashes}{vb1,vb2}
	    \fmf{plain}{ob1,vb2,ob2}
	  \end{fmfgraph*}
	\end{fmffile}
	&  SI\\
	\cline{2-4}
	&$X$&
	\begin{fmffile}{tab-fm3}
	  \rule{0pt}{19pt}
	  \begin{tabular}{m{29pt}m{0pt}m{29pt}}
	    \begin{fmfgraph*}(12,8)
	      \fmfleftn{ic}{2} \fmfrightn{oc}{2}
	      \fmf{plain}{ic1,vc1}
	      \fmf{plain}{vc2,ic2}
	      \fmf{boson,tension=0.5}{vc1,vc2}
	      \fmf{phantom}{oc1,vc1}
	      \fmf{phantom}{vc2,oc2}
	      \fmf{plain,tension=0}{oc1,vc2}
	      \fmf{plain,tension=0}{vc1,oc2}
	    \end{fmfgraph*}
	    &+&
	    \begin{fmfgraph*}(12,8)
	      \fmfleftn{ic}{2} \fmfrightn{oc}{2}
	      \fmf{boson,tension=0.5}{vc1,vc2}
	      \fmf{plain}{ic1,vc1}
	      \fmf{plain}{vc2,ic2}
	      \fmf{plain}{oc1,vc1}
	      \fmf{plain}{vc2,oc2}
	    \end{fmfgraph*}
	  \end{tabular}
	\end{fmffile}
	& SD in chiral limit\\
	\cline{2-4}
	&$\Phi$&
	\begin{fmffile}{tab-fm4}
	  \rule{0pt}{19pt}
	  \begin{tabular}{m{29pt}m{0pt}m{29pt}}
	    \begin{fmfgraph*}(12,8)
	      \fmfleftn{id}{2} \fmfrightn{od}{2}
	      \fmf{plain}{id1,vd1}
	      \fmf{plain}{vd2,id2}
	      \fmf{dashes}{vd1,vd2}
	      \fmf{phantom}{od1,vd1}
	      \fmf{phantom}{vd2,od2}
	      \fmf{plain,tension=0}{od1,vd2}
	      \fmf{plain,tension=0}{vd1,od2}
	    \end{fmfgraph*}
	    &+&
	    \begin{fmfgraph*}(12,8)
	      \fmfleftn{id}{2} \fmfrightn{od}{2}
	      \fmf{plain}{vd1,id1}
	      \fmf{plain}{id2,vd2}
	      \fmf{dashes}{vd1,vd2}
	      \fmf{plain}{od1,vd1}
	      \fmf{plain}{vd2,od2}
	    \end{fmfgraph*}
	  \end{tabular}
	\end{fmffile}
	&  SD  in chiral limit\\
	\hline
	\multirow{2}{*}{Real Vector}
	&$h$
	&
	\rule{0pt}{25pt}
	\begin{fmffile}{tab-v1}
	  \begin{fmfgraph*}(12,8)
	    \fmfleftn{ib}{2} \fmfrightn{ob}{2}
	    \fmf{boson}{ib1,vb1,ib2}
	    \fmf{dashes}{vb1,vb2}
	    \fmf{plain}{ob1,vb2,ob2}
	  \end{fmfgraph*}
	\end{fmffile}
	&  SI\\
	\cline{2-4}
	&$Q$&
	\begin{fmffile}{tab-v2}
	  \rule{0pt}{19pt}
	  \begin{tabular}{m{22pt}m{0pt}m{22pt}}
	    \begin{fmfgraph*}(12,8)
	      \fmfleftn{ic}{2} \fmfrightn{oc}{2}
	      \fmf{boson}{ic1,vc1}
	      \fmf{boson}{ic2,vc2}
	      \fmf{plain}{vc1,vc2}
	      \fmf{plain}{oc1,vc1}
	      \fmf{plain}{vc2,oc2}
	    \end{fmfgraph*}
	    &+&
	    \begin{fmfgraph*}(12,8)
	      \fmfleftn{id}{2} \fmfrightn{od}{2}
	      \fmf{boson}{id1,vd1}
	      \fmf{boson}{vd2,id2}
	      \fmf{plain}{vd2,vd1}
	      \fmf{phantom}{od1,vd1}
	      \fmf{phantom}{vd2,od2}
	      \fmf{plain,tension=0}{od1,vd2}
	      \fmf{plain,tension=0}{vd1,od2}
	    \end{fmfgraph*}
	  \end{tabular}
	\end{fmffile}
	&  SD in chiral limit\\
	\hline
	\multirow{3}{*}{Complex Vector}
	&$Z,Z'$& 
	\rule{0pt}{25pt}
	\begin{fmffile}{tab-v1b}
	  \begin{fmfgraph*}(12,8)
	    \fmfleftn{ia}{2} \fmfrightn{oa}{2}
	    \fmf{boson}{ia1,va1,ia2}
	    \fmf{boson}{va1,va2}
	    \fmf{plain}{oa1,va2,oa2}
	  \end{fmfgraph*}
	\end{fmffile}
	&
	SI\\
	\cline{2-4}
	&$h$
	&
	\rule{0pt}{25pt}
	\begin{fmffile}{tab-v2b}
	  \begin{fmfgraph*}(12,8)
	    \fmfleftn{ib}{2} \fmfrightn{ob}{2}
	    \fmf{boson}{ib1,vb1,ib2}
	    \fmf{dashes}{vb1,vb2}
	    \fmf{plain}{ob1,vb2,ob2}
	  \end{fmfgraph*}
	\end{fmffile}
	&  SI\\
	\cline{2-4}
	&$Q$&
	\rule{0pt}{25pt}
	\begin{fmffile}{tab-v3b}
	  \begin{fmfgraph*}(12,8)
	    \fmfleftn{id}{2} \fmfrightn{od}{2}
	    \fmf{boson}{id1,vd1}
	    \fmf{boson}{id2,vd2}
	    \fmf{plain}{vd1,vd2}
	    \fmf{plain}{od1,vd1}
	    \fmf{plain}{vd2,od2}
	  \end{fmfgraph*}
	  ,
	  \begin{fmfgraph*}(12,8)
	    \fmfleftn{id}{2} \fmfrightn{od}{2}
	    \fmf{boson}{id1,vd1}
	    \fmf{boson}{vd2,id2}
	    \fmf{plain}{vd2,vd1}
	    \fmf{phantom}{od1,vd1}
	    \fmf{phantom}{vd2,od2}
	    \fmf{plain,tension=0}{od1,vd2}
	    \fmf{plain,tension=0}{vd1,od2}
	  \end{fmfgraph*}
	\end{fmffile}
	&  SI, SD\\
	\hline
      \end{tabular}
    \end{center}
    \caption{
    A summary of our results for WIMP-nucleon scattering, for each dark
    matter candidate and mediator. In the Feynman diagrams, scalars are
    represented by dashed lines, fermions by solid lines and vector
    bosons by wavy lines.
    Of the mediators, $h$, $Z'$ and the SM $Z$ are neutral
    under both electromagnetism and color, while $X$, $\Phi$ and $Q$
    transform as triplets under color and carry electric charge.
    }
    \label{tab:spind-summary}
  \end{minipage}
\end{table}

\afterpage{\clearpage}
\subsection{Summary}

Our results are summarised in Table \ref{tab:spind-summary}. We see that the dark 
matter candidates which naturally lead to sizeable spin-dependent WIMP-nucleon 
cross sections without correspondingly large spin-independent cross sections are 
Majorana fermions and real vector bosons. These theories share the feature that 
the dark matter particle is its own anti-particle. We stress however that the 
fact that the dark matter particle is a Majorana fermion or real vector boson is 
not sufficient by itself to guarantee that WIMP-nucleon cross sections are 
primarily spin-dependent. This is only true if the chirality suppressed t-channel 
scalar exchange contribution is either absent or sub-dominant. While this can 
naturally be the case it is certainly not guaranteed.

The interactions of scalars with nucleons are spin-independent. The cross 
sections of Dirac fermions and complex vector bosons in general tend to have a 
sizeable spin-independent component unless the dark sector possesses a discrete 
symmetry similar to charge conjugation, but under which the SM fields are 
invariant. In general incorporating such a symmetry requires complicating the 
theory by adding multiple mediators, except in the case of Dirac fermion dark 
matter interacting through t-channel vector exchange, when very specific choices 
of charge assignments allow such a symmetry to be realized.

What are the constraints from flavor on the class of theories we are 
considering? For WIMP-nucleon scattering mediated by the SM $Z$, flavor 
bounds are automatically satisfied. This also applies to t-channel $Z'$ 
exchange, provided the couplings of the $Z'$ are flavor-diagonal.  
However, couplings of the type quark-WIMP-mediator are very strongly 
constrained by precision flavor experiments, as these couplings will in 
general have a non-trivial flavor structure. In general, we need different 
mediators to couple to up-type and down-type quarks.  Further, in order 
to satisfy these bounds, it may be necessary to introduce either multiple 
flavors of the mediating particle or multiple flavors of the dark matter 
particle, while incorporating some version of a GIM mechanism. These 
considerations, while important, do not impact our conclusions and we 
therefore leave this for future work.

In summary, we see that in each case where the spin-dependent contribution 
naturally dominates, there are either new particles charged under the SM gauge 
groups, or a new $Z'$ gauge boson.  In the next section we investigate the 
implications of this result for the LHC.

\section{Implications for Colliders} 

The analysis above reveals that in the chiral limit there is only one effective 
operator for each of Majorana and real vector boson dark matter which leads to 
spin-dependent interactions with nucleons . Therefore, a spin-dependent 
scattering signal in a direct detection experiment implies a model-independent 
lower bound on the co-efficient of this operator. This in turn places limits on 
the masses of the particles mediating dark matter interactions, which can be 
searched for in collider experiments such as the LHC. For WIMPs whose primary 
interactions with nucleons are spin-dependent, direct detection experiments and 
collider searches are therefore highly correlated. In what follows we consider 
Majorana fermion dark matter and real vector boson dark matter in turn, and 
explore the region of parameter space which is accessible to direct detection 
experiments, and to the LHC.

\subsection{Majorana Fermion Dark Matter}

Starting from the effective Lagrangian
\begin{align}
  L_{\text{eff}} &= 
  d_q\, \bar{\chi} 
  \gamma^\mu\gamma^5
  \chi\;
  \bar{q}
  \gamma_\mu\gamma^5
  q,
\end{align}
we can derive the cross section (see Appendix \ref{sec:scattering}),
\begin{align}
  \sigma_0
  =
  \frac{16 m_{\chi}^2m_N^2}{\pi(m_{\chi}+m_N)^2}
  \left[\sum_{q=u,d,s}{d_q\lambda_q}\right]^2
  J_N(J_N+1)
  .
  \label{eq:maj-cs}
\end{align}

For a free nucleon, $\lambda_q$ is given simply by $\Delta_q^n$, the spin 
fraction of the nucleon carried by quark $q$. $J_N$ is the angular 
momentum of the nucleus, equal to $\frac12$ for free nucleons, while 
$m_{\chi}$ and $m_N$ are the dark matter mass and the mass of the nucleus 
respectively. The quark spin fractions in the proton and the neutron are shown in Table 
\ref{tab:qspin}. 

\begin{figure}[htp]
  \begin{center}
    \hspace{0.1in}
    \subfloat[]{
    \includegraphics{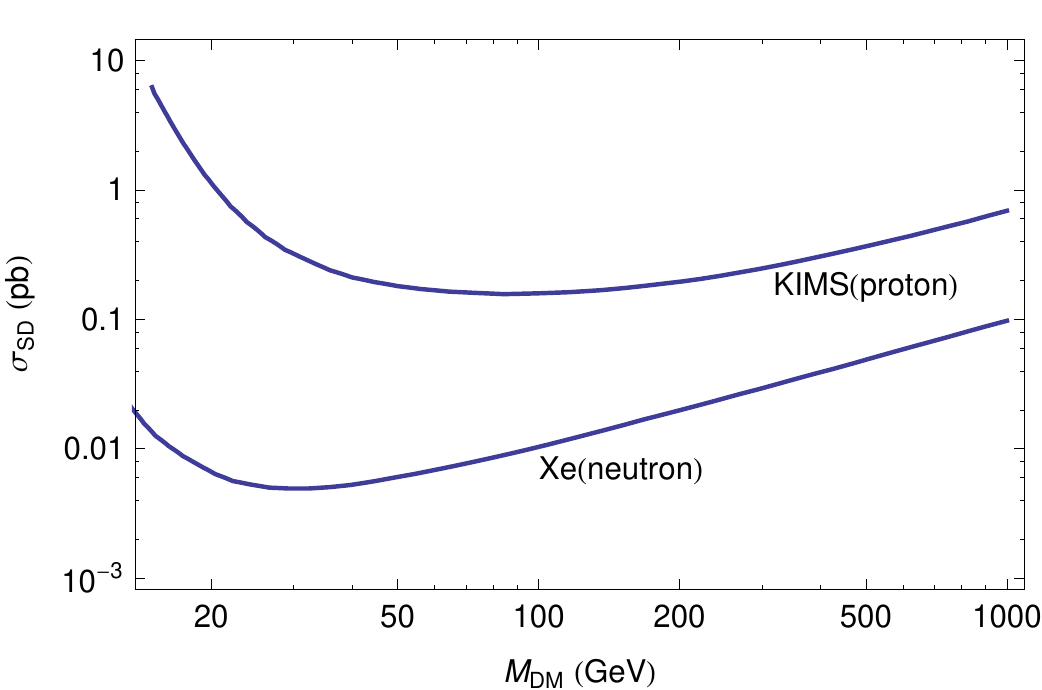}
    \label{fig:sd-direct}
    }
    \vspace{0.3in}
    \\
    \subfloat[]{
    \label{fig:si-direct}
    \includegraphics{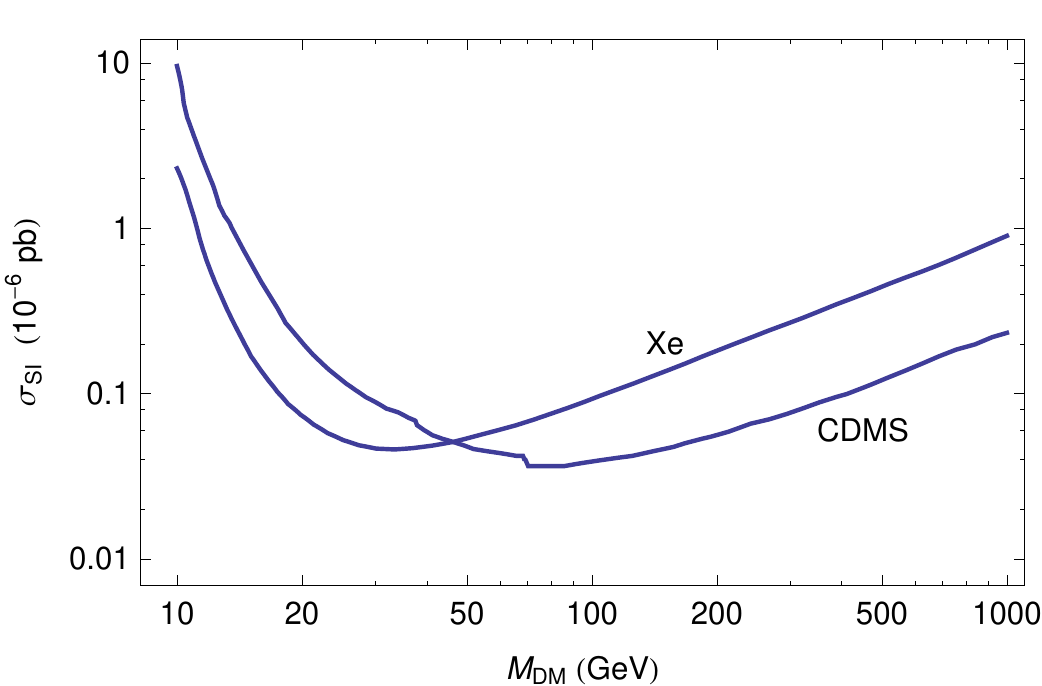}
    }
  \end{center}
  \caption{Current direct detection bounds on the spin-dependent
  (above) and
  spin-independent (below) dark matter-nucleon cross sections. The 
  spin-independent bound assumes that dark matter has equal couplings to 
  protons and neutrons.}
\end{figure}

This cross section is bounded by current experiments. The KIMS 
\cite{Kim:2008zzn} experiment is currently the most sensitive to the 
WIMP-proton coupling while the XENON experiment \cite{Angle:2008we} is the 
most sensitive to the WIMP-neutron coupling. Direct detection experiments 
translate their limits into bounds on the WIMP-nucleon cross section using 
a model-independent framework \cite{Tovey:2000mm}. We see from figure 
\ref{fig:sd-direct} that if the model predicts identical couplings to 
neutrons and protons, then the XENON experiment provides the more 
stringent bound. These experiments will probe new parameter space in the 
near future, with the XENON experiment in particular expected to improve 
the current limit by two orders of magnitude.

\begin{table}
  \begin{center}
  \begin{tabular}{c|cc}
    & proton & neutron\\
    \hline
    $\Delta_u$ &$  0.78\pm0.02$ &$ -0.48\pm0.02$\\
    $\Delta_d$ &$ -0.48\pm0.02$ &$  0.78\pm0.02$\\
    $\Delta_s$ &$ -0.15\pm0.02$ &$ -0.15\pm0.02$\\
  \end{tabular}
  \end{center}
  \caption{Quark spin fractions in the proton and neutron
  \cite{Ellis:2000ds,Mallot:1999qb}}
  \label{tab:qspin}
\end{table}

The values of $d_u$, $d_d$ and $d_s$ which appear in equation 
\eqref{eq:maj-cs} depend on the flavor structure of the theory. As 
explained earlier, dark matter scattering mediated by a $Z'$ is 
automatically consistent with constraints from flavor provided that the 
couplings of the $Z'$ are flavor diagonal. Therefore $d_d = d_s$ in this 
class of models. For theories where WIMP-nucleon scattering is mediated 
by $X_{\mu}$ or $\Phi$, the flavor structure is more complicated. In 
general, a GIM mechanism is required to ensure that flavor bounds are 
satisfied. We therefore allow the possibility of multiple mediators, 
with each mediator associated with a different quark flavor. In general 
the mediators that couple to left- and right-handed quarks are also 
distinct. Flavor constraints are satisfied provided that the mediators 
corresponding to different flavors of quarks are degenerate, and their 
couplings are flavor diagonal.

We now consider in turn the various theories of Majorana fermion dark 
matter which naturally lead to primarily spin-dependent WIMP-nucleon cross 
sections. For each theory we explore the range of dark matter and mediator 
masses which leads to a signal in direct detection experiments, and the
implications for the LHC.
\begin{itemize}
  \item {Via the SM $Z$}\\
    The axial couplings of the SM $Z$ to quarks are proportional to 
    $I_3$, i.e. $d_u=-d_d=-d_s$.
    From figure \ref{fig:zsd} we see that the current spin-independent
    bounds imply that for the spin-dependent
    WIMP-nucleon cross section to be accessible to upcoming experiments
    the dark matter mass must lie below about 400 GeV. Therefore, in most of the parameter space that these 
    experiments 
    will probe the dark matter particle $\chi^0$ and its charged partner 
    $\chi^+$ are kinematically
    accessible to the LHC. Although this result was obtained in the 
    context of a specific benchmark model, we do not expect the results 
    in the general case to to differ significantly.

  \item {Via a $Z'$} \\
    On comparison with equation (\ref{eq:ma}), we see that the 
    value of the coefficient is
    \begin{align}
      d_q &= 
      -\frac{\beta \widetilde{\beta_q}}{m_{Z'}^2}
      .
    \end{align}
    As for the SM $Z$, we set $d_u=-d_d=-d_s$, since this choice is
    naturally consistent with flavor constraints on new physics, and
    yields a conservative estimate for the mediator mass.  The values
    $\beta$ and $\widetilde{\beta}$ equal to a $\frac12$ correspond to
    chiral fermions having unit charge under the $Z'$. We use these as
    representative values. In figure \ref{fig:lhcu} we have shown the
    range of values of the $Z'$ mass that would lead to a signal in
    direct detection experiments at the current bound, or within two
    orders of magnitude of the current bound.  Unfortunately, the
    allowed range of masses is disfavored by precision electroweak
    measurements, direct
    production and four-fermion point interactions~\cite{Appelquist:2002mw,Amsler:2008zzb}, except perhaps
    for very specific charge assignments.

\begin{figure}[h]
  \begin{center}
    \subfloat[]{
    \includegraphics{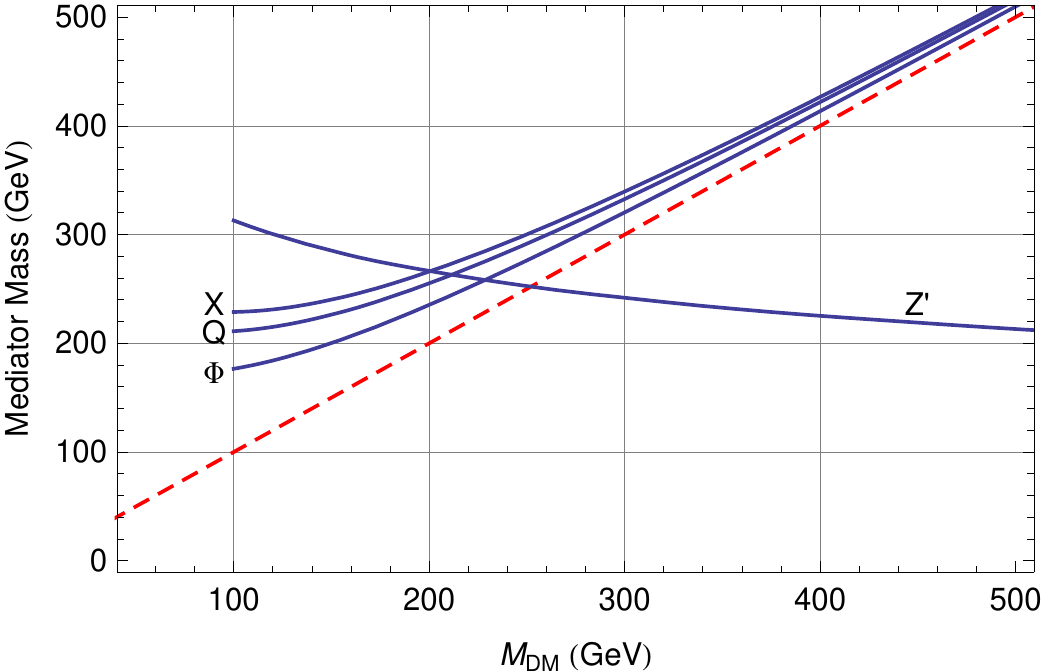}
    }
    \vspace{0.3in}
    \\
    \subfloat[]{
    \includegraphics{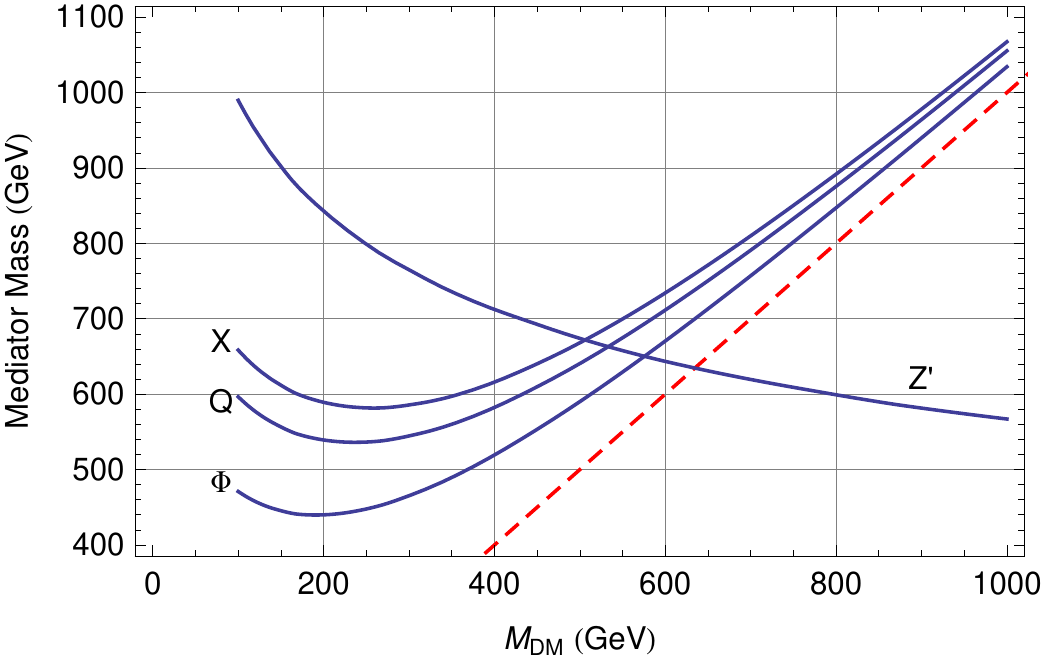}
    }
  \end{center}
  \caption{ Estimates for the mediator masses if direct detection
  experiments see a signal near the present bound (above) or two
  orders of magnitude below the present bound (below).  The colored
  mediators $X$, $\Phi$ and $Q$ are assumed to couple only to up-type
  quarks, while the charges of the $Z'$ are assumed to be proportional
  to the charges of the SM $Z$.  The masses of $X, \Phi$ and $Q$ must
  lie above the red dashed line, which corresponds to where the
  mediator mass is equal to the dark matter mass.  }
  \label{fig:lhcu}
\end{figure}

\begin{figure}[h]
  \begin{center}
    \subfloat[]{
    \includegraphics{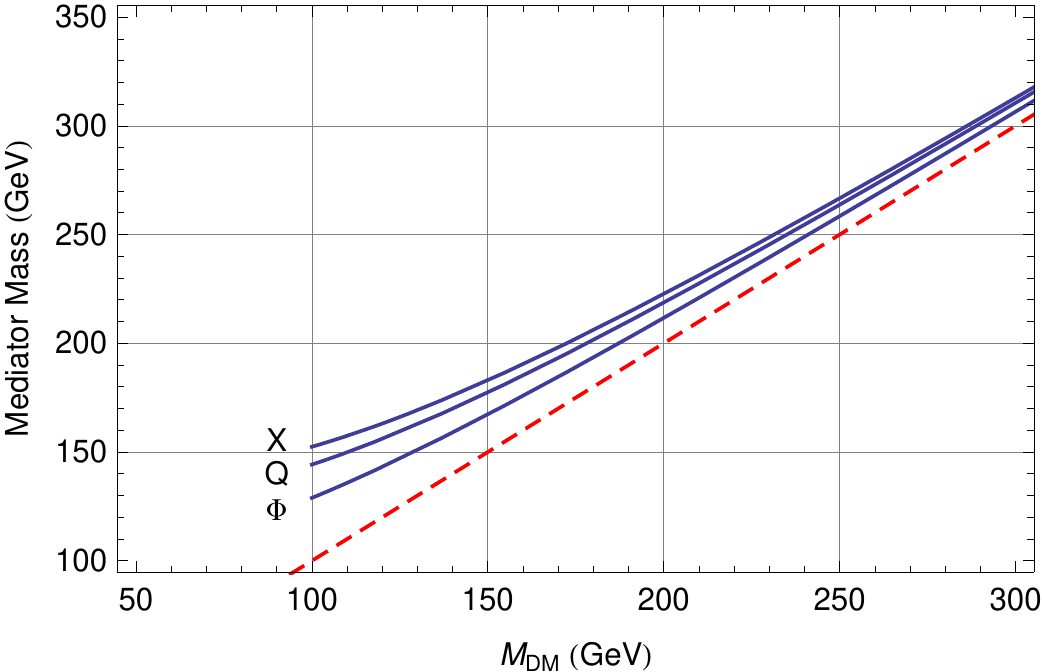}
    }
    \vspace{0.3in}
    \\
    \subfloat[]{
    \includegraphics{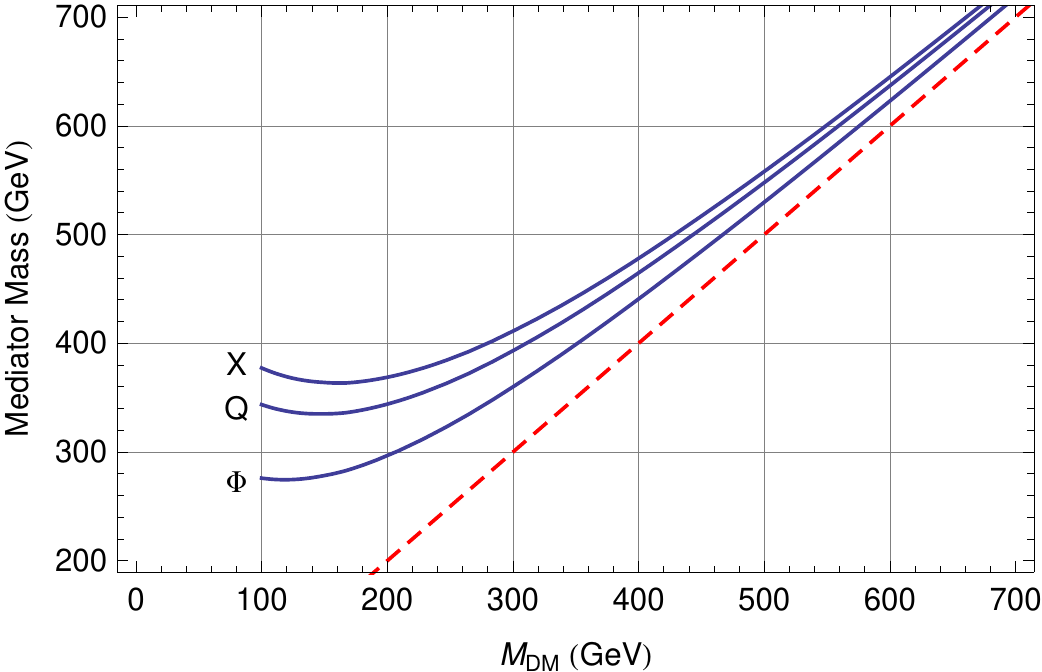}
    }
  \end{center}
  \caption{
  Estimates for the colored mediator masses if direct detection
  experiments see a signal near the present bound (above) or two
  orders of magnitude below the present bound (below), assuming 
  equal couplings to up- and down-type quarks. The signal in
  direct detection experiments is suppressed due to cancellations
  arising from quark spin fractions, leading to lower values of
  the mediator masses.
  }
  \label{fig:lhc}
\end{figure}

\afterpage{\clearpage}
  \item {Via colored vector bosons, $X_{\mu}$}\\
    \begin{align}
      d_q &= 
      -\frac{|\alpha_q|^2 + |\beta_q|^2}{2(m_{X,q}^2-m_{\chi}^2)}
    \end{align}
    In the chiral limit, the $X$ vector bosons which
    couple to left- and right-handed quarks are distinct particles.
    In the absence of tuning, it is therefore natural for either the
    left- or right-handed contribution to dominate. 
    For the left-handed contribution, if the dark matter particle is a
    SM singlet we expect that $d_u = d_d$ as a
    consequence of the SM SU(2$)_{\rm L}$ symmetry. Further, flavor
    constraints require $d_d = d_s$. Then the values of $\Delta_u$,
    $\Delta_d$ and $\Delta_s$ imply that there are large cancellations
    among the contributions of the different left-handed quarks to the
    WIMP-nucleon cross section, which is therefore somewhat
    suppressed. For the right-handed contribution, this cancellation
    can be avoided if $d_u \gg d_d ( = d_s)$, or vice versa. 

    In figures \ref{fig:lhcu} and \ref{fig:lhc} we have plotted the
    range of values of the $X$ masses that would lead to a signal at
    current direct detection experiments. In figure \ref{fig:lhcu} we
    have set $d_d=d_s=0$ and $d_u\neq0$, with $\alpha_u = -\beta_u =
    \frac12$, corresponding to one natural possibility for the
    contribution from right-handed quarks.  In figure \ref{fig:lhc},
    on the other hand, we have set $d_u = d_d = d_s$, with $\alpha_q =
    \beta_q = \frac12$ corresponding to the contribution from
    left-handed quarks.  We see that away from the resonance region at
    $m_{\chi} = m_{X}$, the colored vector boson masses lie at a TeV
    or below in all of parameter space. They are therefore kinematically
    accessible to the LHC, and can be pair-produced through strong
    interactions. The signal is jets + missing energy. Recent 
    model-independent studies of dark matter signals at the LHC
    involving jets + missing energy may be found, for example, in
    \cite{Alwall:2008va,Izaguirre:2010nj}, \cite{Cao:2009uw},
    \cite{Beltran:2010ww}. 

  \item {Via colored scalars, $\Phi$}\\
    \begin{align}
      d_q &= 
      \frac{|\alpha_q|^2 + |\beta_q|^2}{4(m_{\Phi,q}^2-m_\chi^2)}
    \end{align}
    As explained earlier, in the chiral limit the mediators of left- and 
    right-handed interactions are distinct. 
    In figures \ref{fig:lhcu} and \ref{fig:lhc} we have shown the range
    of values of the $\Phi$ mass that would lead to a signal at current
    direct detection experiments. In figure \ref{fig:lhcu}
    we have set $d_u\neq0$ and $d_d=d_s=0$, 
    with $\alpha_u = -\beta_u =\frac12$, corresponding to the right-handed
    contribution. In figure \ref{fig:lhc} we have set
    $d_u = d_d = d_s$ with $\alpha_q = \beta_q =\frac12$, corresponding
    to the left-handed contribution. 
    We see that away from the resonance region the $\Phi$ masses 
    lie below a TeV, which is promising for the LHC. 
\end{itemize}

\subsection{Real Vector Boson Dark Matter}
The effective Lagrangian for WIMP-nucleon scattering in the case 
of real vector boson dark matter takes the form,
\begin{align}
  \mathcal{L}
  &= b_q (\partial_\sigma B_\mu) B_\nu \bar{q} \gamma_\alpha \gamma^5 q
  \epsilon^{\sigma\mu\nu\alpha}
  .
\end{align}
The corresponding cross section is 
\begin{align}
  \sigma_0
  &=
  \frac{8 m_{\chi}^2 m_N^2}{3 \pi \left(m_{\chi}+m_N \right)^2}
  \left[\sum_{q=u,d,s}{b_q\lambda_q}\right]^2 J_N(J_N+1)
  .
\end{align}
Here $b_q$ is related to the mass and couplings of the colored fermions 
$Q$ mediating the interaction,
\begin{align}
  b_q = \frac{|\alpha_q|^2 + |\beta_q|^2}{(m_{Q,q}^2-m_B^2)}
  .
\end{align}
We seek to explore the range of masses of $Q$ and $B_{\mu}$ that give 
rise to a signal at current direct detection experiments, and the 
resulting implications for the LHC. As in the cases of $X$ and $\Phi$, 
in the chiral limit the mediators $Q$ that couple to left-and 
right-handed quarks are in general different particles, and the 
mediators corresponding to different flavors are also distinct. Flavor 
constraints are satisfied provided the mediators associated with 
different flavors are degenerate, and their couplings are flavor 
diagonal. For concreteness we employ exactly the same conventions as 
earlier. Specifically, we first consider $b_u \neq 0$, $b_d = b_s=0$ 
with $\alpha_u = -\beta_u = \frac{1}{2}$, corresponding to the 
contribution from right-handed quarks. The results are plotted in figure 
\ref{fig:lhcu}. We then consider $b_u = b_d = b_s$ with $\alpha_q = 
\beta_q =\frac12$, corresponding to the contribution from left-handed 
quarks. The results are plotted in figure \ref{fig:lhc}. From the 
figures, we see that a signal at current spin-dependent direct detection 
experiments implies that the masses of the colored fermions lie at a TeV 
or below, which is within the kinematic reach of the LHC.

\section{Conclusions}

We have classified dark matter candidates that have the property that 
WIMP-nucleon scattering is dominated by spin-dependent interactions. We have 
established that in this scenario the natural dark matter candidates are Majorana 
fermions or real vector bosons, while scalars are disfavored. Dirac fermion and 
complex vector boson dark matter are also disfavored except for very specific 
choices of quantum numbers.  Therefore spin-dependent WIMP-nucleon cross sections 
are closely associated with theories where the dark matter particle is its own 
anti-particle. Furthermore, we have shown that such theory predicts either new 
particles close to the weak scale with SM quantum numbers, or a new $Z'$ gauge 
boson with mass at or below the TeV scale. In the region of parameter space that 
is of interest to current direct detection experiments, these particles naturally 
lie in a mass range that is kinematically accessible to the Large Hadron Collider 
(LHC).

These results also have implications for experiments involving neutrino 
telescopes searching for the products of dark matter annihilation in the 
sun. The rate of dark matter capture in the sun is controlled by the cross 
section for WIMP-nucleon scattering, which therefore impacts the signal. 
These experiments currently provide stronger limits on the spin-dependent 
WIMP-proton cross section than direct detection 
experiments~\cite{Abbasi:2009uz,Abbasi:2009vg},~\cite{Wikstrom:2009kw}. A 
dark matter signal at neutrino telescopes, if arising from spin-dependent 
WIMP-proton scattering, can be translated into limits on the mediator 
masses. This also has implications for the LHC. We plan to present the results 
of this analysis in a separate publication~\cite{Agrawal:2010ax}.

\acknowledgments
We would like to thank Lian-Tao Wang and Takemichi Okui for useful 
comments. ZC, PA and RKM are supported by the National Science Foundation 
under grant PHY-0801323. CK is supported by the Department of Energy under 
grant DE-FG0296ER50959.

\newpage
\appendix
\appendixpage
\section{Relevant Fierz Identities}
\label{sec:fierz}
We list here the Fierz identities that are necessary to obtain our
results. The results presented are for anti-commuting fields. 
Throughout
these appendices, unless explicitly stated otherwise, we follow the
conventions of Peskin and Schroeder~\cite{Peskin:1995ev}.
\begin{align}
  \bar{q}\chi\bar{\chi}q
  &= -\frac{1}{4} \left[\bar{q}q\bar{\chi}\chi
  +\bar{q}\gamma^{\mu}q\bar{\chi}\gamma_{\mu}\chi
  +\bar{q}\gamma^{5}q\bar{\chi}\gamma^{5}\chi
  -\bar{q}\gamma^{\mu}\gamma^{5}q\bar{\chi}\gamma_{\mu}\gamma^{5}\chi
  +\frac{1}{2}\bar{q}\,\sigma^{\mu\nu}q
  \bar{\chi}\,\sigma_{\mu\nu}\chi\right]
  \\
  \bar{q}\gamma^{5}\chi\bar{\chi}\gamma^{5}q
  &=
  -\frac{1}{4}\left[\bar{q}q\bar{\chi}\chi
  -\bar{q}\gamma^{\mu}q\bar{\chi}\gamma_{\mu}\chi
  +\bar{q}\gamma^{5}q\bar{\chi}\gamma^{5}\chi
  +\bar{q}\gamma^{\mu}\gamma^{5}q\bar{\chi}\gamma_{\mu}\gamma^{5}\chi
  +\frac{1}{2}\bar{q}\,\sigma^{\mu\nu}q
  \bar{\chi}\,\sigma_{\mu\nu}\chi\right]
  \\
  \bar{q}\gamma^{5}\chi\bar{\chi}q
  &=
  -\frac{1}{4}\left[\bar{q}q\bar{\chi}\gamma^{5}\chi
  +\bar{q}\gamma^{\mu}q\bar{\chi}\gamma_{\mu}\gamma^{5}\chi
  +\bar{q}\gamma^{5}q\bar{\chi}\chi
  -\bar{q}\gamma^{\mu}\gamma^{5}q\bar{\chi}\gamma_{\mu}\chi
  -i\epsilon_{\mu\nu\alpha\beta} \bar{q}\,\sigma^{\mu\nu}q
  \bar{\chi}\,\sigma^{\alpha\beta}\chi\right]
  \\
  \bar{q}\gamma^{\mu}\chi\bar{\chi}\gamma_{\mu}q
  &=
  -\left[\bar{q}q\bar{\chi}\chi
  -\frac{1}{2}\bar{q}\gamma^{\mu}q\bar{\chi}\gamma_{\mu}\chi
  -\bar{q}\gamma^{5}q\bar{\chi}\gamma^{5}\chi
  -\frac{1}{2}\bar{q}\gamma^{\mu}\gamma^{5}q
  \bar{\chi}\gamma_{\mu}\gamma^{5}\chi
  \right]
  \\
  \bar{q}\gamma^{\mu}\gamma^{5}\chi 
  \bar{\chi}\gamma_{\mu}\gamma^{5}q
  &=
  -\left[-\bar{q}q\bar{\chi}\chi-
  \frac{1}{2}\bar{q}\gamma^{\mu}q\bar{\chi}\gamma_{\mu}\chi+
  \bar{q}\gamma^{5}q\bar{\chi}\gamma^{5}\chi-
  \frac{1}{2}\bar{q}\gamma^{\mu}\gamma^{5}q
  \bar{\chi}\gamma_{\mu}\gamma^{5}\chi\right]
  \\
  \bar{q}\gamma^{\mu}\chi\bar{\chi}\gamma_{\mu}\gamma^{5}q
  &=
  -\left[
  -\bar{q}q\bar{\chi}\gamma^{5}\chi
  -\frac{1}{2}\bar{q}\gamma^{\mu}q
  \bar{\chi}\gamma_{\mu}\gamma^{5}\chi
  +\bar{q}\gamma^{5}q\bar{\chi}\chi
  -\frac{1}{2}\bar{q}\gamma^{\mu}\gamma^{5}q
  \bar{\chi}\gamma_{\mu}\chi
  \right]
\end{align}

\section{Fermionic dark matter scattering cross section}
\label{sec:scattering}
\label{sec:low-energy}
Let $\chi$ represent a fermionic dark matter particle, $q$ a quark,
$n$ a nucleon and $N$ a nucleus. We consider elastic scattering of the
WIMP off a nucleus
\begin{center}
  $\chi(p_1)+N(p_2) \rightarrow \chi(p_3)+N(p_4)$.
\end{center}

We can write down the effective Lagrangian at the partonic level in
the general case as follows,
\begin{align}
  \mathcal{L}
  &=
  \bar{\chi}(\alpha + \beta \gamma^5)\chi\,
  \bar{q}(\widetilde{\alpha}+\widetilde{\beta} \gamma^5)q
  + \lambda_q\,\bar{\chi}\,\Gamma^{\mu}\chi\,
  \bar{q}\,\widetilde{\Gamma}_{\mu}q
  +c_q\,\bar{\chi}\Lambda^{\mu\nu}\chi\,
  \bar{q}\widetilde{\Lambda}_{\mu\nu}q
  \label{eq:1a}
  .
\end{align}
where the $\mathds{1}$, $\gamma^5$,
$\Gamma$ and $\Lambda$ matrices span the Dirac matrix subspace.

The average speed of $\chi$ in the halo is $300$ km/s. This means that the 
scattering with nuclei in direct detection experiments is highly 
non-relativistic and calls for an appropriate treatment. In particular, 
cross sections which are velocity suppressed are generally smaller by a 
factor of order $10^{-6}$, and can be neglected.

It is important to identify the terms which are relevant
in the low-energy limit. This can be done schematically as follows.
We can expand a fermion field $\psi \sim
au+b^{\dagger} v$ where $u$ and $v$ are independent spinor solutions to
Dirac equation. In the non-relativistic limit, $u$, $v$ reduce to
\begin{align}
  u&=
  \begin{pmatrix}
    \sqrt{p.\sigma} \xi \\
    \sqrt{p.\bar{\sigma}} \xi
  \end{pmatrix}
  \xrightarrow{\text{{NR limit}}}
  \sqrt{m}
  \begin{pmatrix}
    \xi \\
    \xi
  \end{pmatrix}
  & v
  &=
  \begin{pmatrix}
    \sqrt{p.\sigma}\eta \\
    -\sqrt{p.\bar{\sigma}} \eta
  \end{pmatrix}
  \xrightarrow{\text{{NR limit}}}
  \sqrt{m}
  \begin{pmatrix}
    \eta\\
    -\eta
  \end{pmatrix}
  .
\end{align}
From this it follows that,
\begin{align}
  \bar{\psi}\psi
  &\approx
  2m\left[a^{\dagger}a+b^{\dagger}b\right]
  \label{eq:sleb}
  \\
  \bar{\psi}\gamma^{5}\psi
  &\approx
  0
  \\
  \bar{\psi}\gamma^{\mu}\psi
  &\approx
  2m\left[a^{\dagger}a+bb^{\dagger}\right]
  \delta^{\mu0}
  \\
  \bar{\psi}\gamma^{\mu}\gamma^{5}\psi
  &\approx
  2m\left[a^{\dagger}a
  \left(\xi^{\dagger}\sigma^{i}\xi\right)
  +bb^{\dagger}\left(\eta^{\dagger}\sigma^{i}
  \eta\right)\right]\delta^{\mu i}
  \\
  \bar{\psi}\,\sigma^{\mu\nu}\psi
  &\approx
  2m\left[
  a^{\dagger}a
  \left(\xi^{\dagger}
  \sigma^{k}
  \xi\right)+
  b^{\dagger}b\left(\eta^{\dagger}
  \sigma^{k}\eta\right)\right]
  \delta^{\mu i}\delta^{\nu j}
  \epsilon^{ijk}
  .
\end{align}
We note the following features.
\begin{enumerate}
  \item The scalar and vector bilinears just yield the number
    operator. Therefore, when these operators are evaluated in the nuclear
    state, they add coherently, generating spin-independent interactions.
    The axial-vector and the tensor bilinears on the other hand yield
    the spin-operator, and hence couple to the net spin of the
    nucleus.
  \item The vector interaction picks out the temporal component, while the
    axial-vector interaction picks out the spatial component. Therefore,
    4-fermion operators mixing these will be velocity suppressed.
\end{enumerate}

This analysis is schematic and does not account for the complications
arising from strong nuclear dynamics in the case of quark bilinears.
However, as we shall see later, incorporating these effects does not
alter these conclusions. From this it follows that only the following
terms survive in the non-relativistic regime:
\begin{align}
  \mathcal{L}
  &=
  a_q\,\bar{\chi}\chi\,
  \bar{q}q
  +b_q\,\bar{\chi}\gamma^{\mu}\chi\,
  \bar{q}\gamma_{\mu}q
  +d_q\,\bar{\chi}\gamma^{\mu}\gamma^5\chi\,
  \bar{q}\gamma_{\mu}\gamma^5q
  +c_q\,\bar{\chi}\,\sigma^{\mu\nu}\chi\,
  \bar{q}\,\sigma_{\mu\nu}q.
  \label{eq:1b}
\end{align}
 The first two terms lead to spin-independent scattering, while the 
remaining
two lead to spin-dependent scattering. The particle $\chi$ could be either
a Dirac or a Majorana fermion. We specify the distinction where
applicable.

Since the momentum transfers in WIMP-nucleus scattering are generally small
compared to the characteristic nuclear scales, dark matter cross sections
are expressed in terms of the cross section at zero momentum transfer,
$\sigma_0$. This is itself a good approximation to the total cross section
$\sigma$ if the nucleus is small. For scattering off larger nuclei, the
momentum-transfer dependence of the cross section can be parametrized into
a form factor $F(|\vec{q}|)$ \cite{Jungman:1995df}. 
Define,
\begin{align}
  \sigma_0
  &=
  \int_0^{4 \mu^2 v^2}
  d|\vec{q}|^2\;
  \frac{d\sigma(q=0)}{d|\vec{q}|^2}
  .
\end{align}
where $\mu$ is the reduced mass, and $v$ is the velocity.
The actual cross section can be written in terms of $\sigma_0$,
\begin{align}
  \sigma
  &=
  \int d|\vec{q}|^2\;
  \frac{d\sigma}{d|\vec{q}|^2}
  \\&=
  \int d|\vec{q}|^2 \;
  F^2(|\vec{q}|)\,
  \frac{d\sigma(q=0)}{d|\vec{q}|^2}
  \\&=
  \frac{\sigma_0}{4\mu^2v^2}
  \int d|\vec{q}|^2 \;
  F^2(|\vec{q}|)\,
  .
\end{align}
It is conventional to work with $\sigma_0$ to calculate event rates,
since the form factor integral is independent of the details of the
particle physics model (it only depends on the mass of the dark matter
particle). The form factor obviously satisfies $F(0) = 1$. Numerical
or analytical estimates for form factors are available
\cite{Gould1992,Jungman:1995df}.

\subsection{Scalar Interaction}
We begin with the Lagrangian,
\begin{align}
  \mathcal{L}
  &=
  a_q\,\bar{\chi}\chi\,
  \bar{q}q.
\end{align}
Dirac and Majorana particles can both have scalar interactions.  We
assume a Dirac particle to start with.  We work out the amplitude in
detail in this case to make the definitions of the various matrix elements
clear.  The S-matrix element arising from this interaction has the form
\begin{align}
  \mathcal{M}_{if}\;
  \delta^{(4)}
  (p_1+p_2-p_3-p_4)
  &=
  \sum_{q}
  \int
  d^4 x\;
  a_q \langle \chi_f,N_f|
  \bar{\chi}(x) \chi(x)\,
  \bar{q}(x)q(x)
  |\chi_i,N_i\rangle 
  .
\end{align}
We can separate the co-ordinate dependence using the translation
operator. Consider the matrix element,
\begin{align}
  \langle N_f({\bf p_4})|\,
  \bar{q}(x)q(x)
  |N_i({\bf p_2})\rangle
  &=
  \langle N_f({\bf p_4})|\,
  e^{iP\cdot x}
  \bar{q}(0)
  e^{-iP\cdot x}
  e^{iP\cdot x}
  q(0)
  e^{-iP\cdot x}
  |N_i({\bf p_2})\rangle
  \nonumber
  \\&=
  \langle N_f({\bf p_4})|\,
  \bar{q}(0)
  q(0)
  |N_i({\bf p_2})\rangle
  e^{-i(p_2-p_4)\cdot x}
  .
\end{align}
In the limit of small momentum transfer this becomes
\begin{align}
\langle N_f({\bf p_4})|\,
  \bar{q}(x)q(x)
  |N_i({\bf p_2})\rangle
&\approx
  \langle N_f|\,
  \bar{q}q\,
  |N_i\rangle \;
  e^{-i(p_2-p_4)\cdot x}
  .
\end{align}
where $|N_i\rangle $ represents a state corresponding to a nucleus at
rest.
Therefore, in the low energy limit, we can write the amplitude in
terms of the quark matrix elements in nuclear states at rest. These
matrix elements here on are position independent low-energy
quantities. The amplitude is then
\begin{align}
  \mathcal{M}_{if}\;
  \delta^{(4)}
  (p_1+p_2-p_3-p_4)
  &=
  \sum_{q}
  \int
  d^4 x\;
  a_q \,
  \langle N_f|
  \,\bar{q}q \,
  |N_i\rangle \;
  \langle \chi_f|\,
  \bar{\chi} \chi\,
  |\chi_i\rangle \;
  e^{-i(p_1+p_2-p_3-p_4).x}
  \nonumber
  \\&=
  \sum_{q}
  a_q \;
  \langle N_f|
  \,\bar{q}q \,
  |N_i\rangle \;
  \langle \chi_f|\,
  \bar{\chi} \chi\,
  |\chi_i\rangle \;
  \delta^{(4)}(p_1+p_2-p_3-p_4)
  .
\end{align}
Therefore,
\begin{align}
  \mathcal{M}_{if}
  &=
  \sum_{q}
  a_q
  \langle N_f|
  \bar{q}q
  |N_i\rangle 
  \langle \chi_f|
  \bar{\chi} \chi\,
  |\chi_i\rangle 
  .
\end{align}
In this expression we are employing the conventional relativistic
normalization for the one particle states,
\begin{align}
  \langle N({\bf p}) | N({\bf q})\rangle
  = 2E_{\bf p} \delta^{(3)}(\bf p-q)
  .
\end{align}
On the other hand, the nuclear physics matrix elements we seek to
determine are generally expressed in terms of states normalized
according to the non-relativistic convention,
\begin{align}
  \langle \widetilde N({\bf p}) | \widetilde N({\bf q})\rangle
  = \delta^{(3)}(\bf p-q)
  \label{eq:nrnorm}
  .
\end{align}
Therefore, in the non-relativistic convention,
\begin{align}
  \mathcal{M}_{if}
  &=
  4m_{\chi}m_N
  \sum_{q}
  a_q\langle\widetilde N_f|\,
  \bar{q}q\,
  |\widetilde N_i\rangle
  \label{eq:Minnr}
  .
\end{align}
In equation \eqref{eq:Minnr}, we use the low-energy expression 
(equation \eqref{eq:sleb}) for $\chi$, which gives us a factor of
$2m_\chi$. The other factor of $2 m_N$ comes from the relative factor
in relativistic normalization. In the case where $\chi$ is Majorana,
both the $a^\dagger a$ and $b^\dagger b$ terms would contribute,
giving an additional factor of two over equation \eqref{eq:Minnr}. The
nuclear states $|\widetilde N\rangle$ are now normalized
non-relativistically.  We must now evaluate the quark operator matrix
element in the nuclear state.

The matrix element of the light quarks $(q=u,d,s)$ in the neutron and
proton can be computed in chiral perturbation theory from
measurements of pion-nucleon sigma term
\cite{Cheng:1988cz,HaiYangCheng1989347,Gasser1991252}.
\begin{align}
  \langle n|\,m_q\, \bar{q}q\,|n\rangle  &= m_n f^{(n)}_{Tq}
  .
\end{align}
where $n$ represents either the proton or the neutron.

The heavy quarks contribute to the mass of the nucleon through the
triangle diagram \cite{Shifman:1978zn}. Using the heavy quark expansion,
it can be shown that the matrix element for heavy quarks is,
\begin{align}
  \langle n|m_q\, \bar{q}q|n\rangle
  &= \frac{2}{27}m_n
  \left(1-\sum_{q=u,d,s}f^{(n)}_{Tq}\right)
  .
\end{align}
for $q=c,b,t$.
Thus, we can define effective coupling of the dark matter with
protons as
\begin{align}
  \frac{f_p}{m_p} &=
  \sum_{q=u,d,s}
  a_q\frac{f^{(p)}_{Tq}}{m_q}
  +\frac{2}{27}
  \left(1-\sum_{q=u,d,s}f^{(p)}_{Tq}\right)
  \sum_{q=c,b,t}\frac{a_q}{m_q}.
\end{align}
An analogous expression holds for the coupling to neutrons.  We use
$f_{Tu}^{(p)}=0.020\pm0.004$, $f_{Td}^{(p)}=0.026\pm0.005$,
$f_{Tu}^{(n)}=0.014\pm0.003$, $f_{Td}^{(n)}=0.036\pm0.008$,  and
$f_{Ts}^{(p,n)}=0.118\pm0.062$~\cite{Ellis:2000ds}.

The scalar interaction couples left and right-handed quarks.
Therefore, we expect the coupling $a_q$ to be proportional to the mass
of the quarks (unless there are additional sources of chiral symmetry
breaking in the theory).  Therefore, the ratio $a_q/m_q$ is generally
not large even for small quark masses.

Performing the sum over the entire nucleus gives us the following
expression for the S-matrix element,
\begin{align}
  \mathcal{M}_{if} &=
  4m_{\chi}m_N
  \left[
  Z f_p + (A-Z)f_n
  \right].
\end{align}
This leads to the following expression for the cross section of a Dirac
dark matter particle at zero-momentum transfer,
\begin{align}
  \sigma_0 = \frac{\mu^2}{\pi}\left[
  Z f_p + (A-Z)f_n
  \right]^2.
\end{align}
where $\mu$ is the reduced mass of the WIMP-nucleus system.

The only difference for a Majorana particle in the calculation leading up
to here is the factor of two noted earlier. Therefore, the
corresponding cross section for a Majorana particle is simply
\begin{align}
  \sigma_0 = \frac{4\mu^2}{\pi}\left[
  Z f_p + (A-Z)f_n
  \right]^2
  .
\end{align}

\subsection{Vector Interaction}
The vector interaction also contributes to the spin-independent
coupling. The Majorana fermion does not couple to the vector current,
so the following discussion applies only to a Dirac fermion.
\begin{align}
  \mathcal{L}
  &=
  b_q\,\bar{\chi}\gamma^{\mu}\chi\,
  \bar{q}\gamma_{\mu}q
  .
  \label{eq:1c}
\end{align}
The calculation proceeds as in the previous case. Writing the matrix
element in the low-energy limit,
\begin{align}
  \mathcal{M}_{if}
  &=\
  \sum_{q}
  b_q
  \langle \chi_f|
  \bar{\chi}\gamma^\mu\chi\,
  |\chi_i\rangle 
  \langle N_f|
  \bar{q}\gamma_\mu q
  |N_i\rangle \\
  &=
  4m_{\chi}m_N
  \sum_{q}
  b_q\langle \widetilde N_f|
  \bar{q}\gamma^\mu q
  |\widetilde N_i\rangle .
  \label{eq:5b}
\end{align}
The sea-quarks and the gluons do not contribute to the vector current.
The valence quark contributions all add up due to the conservation of
the vector current. Therefore, the coupling to protons and neutrons is
now simply given as,
\begin{align}
  b_p &= 2b_u + b_d\\
  b_n &= b_u + 2b_d.
\end{align}
When the sum over the entire nucleus is performed, we get a form
very similar to the scalar case considered earlier,
\begin{align}
  \mathcal{M}_{if} &=
  4m_{\chi}m_N
  \left[
  Z b_p + (A-Z)b_n
  \right].
\end{align}
This leads to the cross section
\begin{align}
  \sigma_0 = \frac{\mu^2}{\pi}\left[
  Z b_p + (A-Z)b_n
  \right]^2.
\end{align}

\subsection{Axial-Vector Interaction}
The axial-vector interaction will be seen to be spin-dependent. The
Lagrangian is
\begin{align}
  \mathcal{L}
  &=
  d_q\,\bar{\chi}\gamma^{\mu}\gamma^5\chi\,
  \bar{q}\gamma_{\mu}\gamma^5q.
\end{align}
This leads to the following matrix element in the limit of zero momentum
transfer
\begin{align}
  \mathcal{M}_{if}
  &=
  \sum_{q=u,d,s}
  d_q
  \langle \chi_f |
  \bar{\chi}\gamma^{\mu}\gamma^5\chi
  |\chi_i\rangle
  \langle N_f |
  \bar{q}\gamma_{\mu}\gamma^5 q
  |N_i\rangle.
  \label{eq:Maxial}
\end{align}
The sum is only over the light quarks because heavy quarks do not
contribute significantly to the spins of neutrons or protons.  As before
we go to the non-relativistic normalization,
\begin{align}
  \mathcal{M}_{if}
  &=
  4 m_\chi m_N
  \sum_{q=u,d,s}
  d_q
  \langle \widetilde \chi_f |
  \bar{\chi}\gamma^{\mu}\gamma^5\chi
  |\widetilde \chi_i\rangle
  \langle \widetilde N_f |
  \bar{q}\gamma_{\mu}\gamma^5 q
  |\widetilde N_i\rangle.
\end{align}
Recalling the expressions for the spinors in the non-relativistic
limit,
\begin{align}
  \langle \widetilde \chi_f |
  \bar{\chi}\gamma^{\mu}\gamma^5\chi
  |\widetilde \chi_i\rangle
  &=
  2\langle \widetilde \chi_f |
  (S_{\chi})_i
  |\widetilde \chi_i\rangle
  \delta_i^\mu.
\end{align}
We can write the quark spin operator in terms of the spin expectation
values of the proton and the neutron in the nucleus,
\begin{align}
  \langle \widetilde N_f |
  \bar{q}\gamma^{\mu}\gamma^5 q
  |\widetilde N_i\rangle
  &=
  2 \delta^\mu_i 
  \left(
  \langle \widetilde N_f |
  (S_{p})_i \Delta_q^p
  |\widetilde N_i\rangle
  +
  \langle \widetilde N_f |
  (S_{n})_i \Delta_q^n
  |\widetilde N_i\rangle
  \right).
\end{align}
where $\Delta_q^n$ is the part of the spin of the nucleon $n$ carried by
quark $q$. 
The nuclear state is specified by the angular momentum quantum numbers
($J_N, J_{Nz}$). Using the Wigner-Eckart theorem, we can write the
combination of the spin-operators above in terms of the nuclear spin
operator \cite{Cheng:2002ej}.
\begin{align}
  \langle \vec{S}_{p}\rangle \Delta_q^p
  +
  \langle \vec{S}_{n}\rangle \Delta_q^n
  \equiv
  \lambda_q\langle \wt N_f|
  \vec{J}_N|\wt N_i\rangle.
  \label{eq:6}
\end{align}
Conventionally, the angular momentum matrix elements are reported in the
$z$-projection in the highest $M_J$ state~\cite{Ressell:1993qm}.
\begin{align}
  \langle S \rangle 
  \equiv
  \langle J,M_J=J | S_z |J, M_J=J\rangle .
\end{align}
Then, the constant of proportionality is given by
\begin{align}
  \lambda_q
  &=
  \frac{\langle {S}_{p}\rangle}{J_N} \Delta_q^p
  +
  \frac{\langle {S}_{n}\rangle}{J_N}\Delta_q^n,
  \label{eq:6b}
\end{align}
where $\langle S_{p,n}\rangle /J_N$ is the fraction of the nuclear
spin carried by protons or neutrons. For example, we can estimate
$\lambda_q$ in single-particle shell model of nuclei
\cite{Goodman:1984dc,Engel:1991wq}.  Here the spin of the nucleus is
due to the unpaired nucleon $n$. Then
\begin{align}
  \lambda_q
  &=
  \Delta_q^n
  \frac{\langle \wt{N}_f|
  \vec{S}_n\cdot\vec{J}_N
  |\wt{N}_i\rangle }
  {J_N(J_N+1)}
  =\frac{1}{2}
  \Delta_q^n
  \left[1-\frac{L_n(L_n+1)-S_n(S_n+1)}{J_N(J_N+1)}\right].
  \label{eq:7}
\end{align}
However, the value of $\lambda_q$ is different in cases when
the shell-model fails, and must then be estimated numerically.

We can now write the scattering amplitude in terms of nuclear spin,
\begin{align}
  \mathcal{M}_{if}
  &=
  16\,m_{\chi}m_N
  \sum_{q=u,d,s}{d_q\lambda_q}
  \langle \wt N_f|\vec{J}_N|\wt N_i\rangle
  \cdot\langle \wt\chi_f|\vec{S}_{\chi}|\wt\chi_i\rangle.
\end{align}
Squaring and summing/averaging over final and initial states,
\begin{align}
  \langle |\mathcal{M}|^2\rangle 
  &=
  \frac{1}{2\left(2J_N+1\right)}
  \sum_{\chi_i,\chi_f,N_i,N_f}
  {\left|\mathcal{M}_{if}\right|^2}
  \nonumber
  \\
  &=
  \frac{256 m_{\chi}^2m_N^2}{2(2J_N+1)}
  \left[\sum_{q=u,d,s}{d_q\lambda_q}\right]^2
  \frac12 \sum_{N_f}
  \langle \wt N_f|J_N^2|\wt N_f\rangle
  \nonumber
  \\
  &=
  64m_{\chi}^2m_N^2
  \left[\sum_{q=u,d,s}
  {d_q\lambda_q}
  \right]^2
  J_N(J_N+1).
  \label{eq:8}
\end{align}
Hence the cross section in the NR limit is given as
\begin{align}
  \sigma_0
  =
  \frac{4\mu^2}{\pi}
  \left[\sum_{q=u,d,s}{d_q\lambda q}\right]^2
  J_N(J_N+1).
  \label{eq:9}
\end{align}
We can repeat this exercise for the Majorana fermion, with the only
difference again being a factor of two in the low energy expression
for the bilinear. Therefore, the expression for cross section for the
Majorana particle is just 4 times the cross section for a Dirac
particle.
\begin{align}
  \sigma_0
  =
  \frac{16 \mu^2}{\pi}
  \left[\sum_{q=u,d,s}{d_q\lambda_q}\right]^2
  J_N(J_N+1).
  \label{eq:9b}
\end{align}
Values of $d_q$ are obtained from theory while the value $\lambda_q$
depends on the nucleus. For scattering off free protons (neutrons),
$\lambda_q$ reduces to $\Delta_q^p$ ($\Delta_q^n$).

\subsection{Tensor Interaction}
We now consider the tensor interaction
\begin{align}
  \mathcal{L}
  =\sum_{q}{b_q\;\bar{\chi}\,\sigma^{\mu\nu}\chi
  \;\bar{q}\,\sigma_{\mu\nu}q}.
  \label{eq:13}
\end{align}
In the non-relativistic limit, this interaction will also yield
spin-dependent interactions.  Again, the current in the case of a
Majorana fermion vanishes, so the following applies to a Dirac fermion
only. We note that in the non-relativistic limit the bilinears that
arise are very similar to the axial-vector case. Therefore, we can
adapt that calculation to this after accounting for some extra
factors.  Specifically
\begin{align}
  \mathcal{M}_{tensor}
  &=
  \sum_{q=u,d,s}
  b_q\,
  \langle \chi_f, N_f|\,
  {\bar{\chi}\,\sigma^{\mu\nu}\chi\,
  \bar{q}\,\sigma_{\mu\nu}q}\,
  |\chi_i,N_i\rangle
  \\ &=
  2\,\mathcal{M}_{\text{axial-vector}}.
  \label{eq:17}
\end{align}
Since everything else including the kinematic factors are the same,
this simply translates into a factor of 4 in the cross section.
Thus,
\begin{align}
  \sigma_0
  =
  \frac{16 \mu^2}{\pi}
  \left(\sum_{q=u,d,s}{b_q\lambda_q}\right)^2
  J_N(J_N+1).
  \label{eq:9c}
\end{align}

\section{Vector dark matter scattering cross section}

We saw that in the case of real vector boson dark matter the effective 
interactions that survive in the non-relativistic limit have either the scalar or 
the axial-vector form. Using the analysis above, we can easily calculate the 
cross sections in these cases.

\subsection{Scalar Interaction}
The scalar term in the Lagrangian is written as
\begin{align}
  \mathcal{L}
  &= a_q m_B B_\mu B^\mu \bar{q} q.
\end{align}
The additional factor of mass in the definition is simply to maintain the
analogy with the four-fermion interaction.
Then, in the limit of zero momentum transfer
\begin{align}
  \mathcal{M}_{if}
  &= 4 m_N\;m_B\, [Z f_p + (A-Z) f_n]
  \epsilon^{\mu*}(p_3) \epsilon_\mu(p_1).
\end{align}
where $\vec{p}_3 \approx \vec{p}_1 \approx 0$. The quantities $f_p$
and $f_n$ are
defined exactly as in the scalar case.
\begin{align}
  \frac{f_p}{m_p} &=
  \sum_{q=u,d,s}
  a_q\frac{f^{(p)}_{Tq}}{m_q}
  +\frac{2}{27}
  \left(1-\sum_{q=u,d,s}f^{(p)}_{Tq}\right)
  \sum_{q=c,b,t}\frac{a_q}{m_q}.
\end{align}
This leads to the cross section,
\begin{align}
  \sigma_0
  &=
  \frac{\mu^2}{\pi}
  [Z f_p + (A-Z) f_n]^2,
\end{align}
where $\mu$ is the reduced mass.

\subsection{Axial-Vector Interaction}
The form of the Lagrangian is
\begin{align}
  \mathcal{L}
  &= b_q (\partial_\sigma B^\mu) B^\nu \bar{q} \gamma_\alpha \gamma^5 q
  \epsilon^{\sigma\mu\nu\alpha}.
\end{align}
This leads to the matrix element
\begin{align}
  \mathcal{M}_{if}
  &= 4 b_q m_N m_B\,
  \epsilon^{0\mu\nu\alpha}
  \epsilon_{\mu}^*(p_3) \epsilon_\nu(p_1)
  \langle \wt N_f | \bar{q}\gamma_\alpha\gamma^5 q|\wt N_i\rangle,
\end{align}
where as before $\vec{p}_3 \approx \vec{p}_1 \approx 0$.
\begin{align}
  \langle|\mathcal{M} |^2\rangle
  &= \frac{128}{3(2J_N+1)} m_N^2 m_B^2\,
  \left[\sum_{q=u,d,s}{b_q\lambda_q}\right]^2
  \sum_{N_f}
  \langle \wt N_f | 
  J_N^2|
  \wt N_f\rangle.
\end{align}
The corresponding cross section is then
\begin{align}
  \sigma_0
  &=
  \frac{8\mu^2}{3\pi}
  \left[\sum_{q=u,d,s}{b_q\lambda_q}\right]^2
  J_N(J_N+1),
\end{align}
where $\mu$ is again the reduced mass of the WIMP-nucleus system.

\bibliography{references}
\bibliographystyle{JHEP}    

\end{document}